\documentclass[preprint,aps]{revtex4}
\usepackage{graphicx}
\usepackage{setspace}
\usepackage{amsmath}
\usepackage{amssymb}

\begin{document}

\title{A search for periodic modulations of the solar neutrino flux in Super-Kamiokande-I}

\newcounter{foots}
\newcounter{notes}
\newcommand{\authoraticrr}{$^{a}$}
\newcommand{\authoratncen}{$^{b}$}
\newcommand{\authoratbu}{$^{c}$}
\newcommand{\authoratbnl}{$^{d}$}
\newcommand{\authoratuci}{$^{e}$}
\newcommand{\authoratcsu}{$^{f}$}
\newcommand{\authoratcnu}{$^{g}$}
\newcommand{\authoratgmu}{$^{h}$}
\newcommand{\authoratgifu}{$^{i}$}
\newcommand{\authoratuh}{$^{j}$}
\newcommand{\authoratkek}{$^{k}$}
\newcommand{\authoratkobe}{$^{l}$}
\newcommand{\authoratkyoto}{$^{m}$}
\newcommand{\authoratlanl}{$^{n}$}
\newcommand{\authoratlsu}{$^{o}$}
\newcommand{\authoratumd}{$^{p}$}
\newcommand{\authoratmit}{$^{q}$}
\newcommand{\authoratduluth}{$^{r}$}
\newcommand{\authoratsuny}{$^{s}$}
\newcommand{\authoratnagoya}{$^{t}$}
\newcommand{\authoratniigata}{$^{u}$}
\newcommand{\authoratosaka}{$^{v}$}
\newcommand{\authoratseoul}{$^{w}$}
\newcommand{\authoratshizuokaseika}{$^{x}$}
\newcommand{\authoratshizuoka}{$^{y}$}
\newcommand{\authoratskku}{$^{z}$}
\newcommand{\authorattohoku}{$^{aa}$}
\newcommand{\authorattokyo}{$^{bb}$}
\newcommand{\authorattokai}{$^{cc}$}
\newcommand{\authorattit}{$^{dd}$}
\newcommand{\authoratwarsaw}{$^{ee}$}
\newcommand{\authoratuw}{$^{ff}$}

\newcommand{\addressoficrr}[1]{$^{a}$ #1 }
\newcommand{\addressofncen}[1]{$^{b}$ #1 }
\newcommand{\addressofbu}[1]{$^{c}$ #1 }
\newcommand{\addressofbnl}[1]{$^{d}$ #1 }
\newcommand{\addressofuci}[1]{$^{e}$ #1 }
\newcommand{\addressofcnu}[1]{$^{f}$ #1 }
\newcommand{\addressofcsu}[1]{$^{g}$ #1 }
\newcommand{\addressofgmu}[1]{$^{h}$ #1 }
\newcommand{\addressofgifu}[1]{$^{i}$ #1 }
\newcommand{\addressofuh}[1]{$^{j}$ #1 }
\newcommand{\addressofkek}[1]{$^{k}$ #1 }
\newcommand{\addressofkobe}[1]{$^{l}$ #1 }
\newcommand{\addressofkyoto}[1]{$^{m}$ #1 }
\newcommand{\addressoflanl}[1]{$^{n}$ #1 }
\newcommand{\addressoflsu}[1]{$^{o}$ #1 }
\newcommand{\addressofumd}[1]{$^{p}$ #1 }
\newcommand{\addressofmit}[1]{$^{q}$ #1 }
\newcommand{\addressofduluth}[1]{$^{r}$ #1 }
\newcommand{\addressofsuny}[1]{$^{s}$ #1 }
\newcommand{\addressofnagoya}[1]{$^{t}$ #1 }
\newcommand{\addressofniigata}[1]{$^{u}$ #1 }
\newcommand{\addressofosaka}[1]{$^{v}$ #1 }
\newcommand{\addressofseoul}[1]{$^{w}$ #1 }
\newcommand{\addressofshizuokaseika}[1]{$^{x}$ #1 }
\newcommand{\addressofshizuoka}[1]{$^{y}$ #1 }
\newcommand{\addressofskku}[1]{$^{z}$ #1 }
\newcommand{\addressoftohoku}[1]{$^{aa}$ #1 }
\newcommand{\addressoftokyo}[1]{$^{bb}$ #1 }
\newcommand{\addressoftokai}[1]{$^{cc}$ #1 }
\newcommand{\addressoftit}[1]{$^{dd}$ #1 }
\newcommand{\addressofwarsaw}[1]{$^{ee}$ #1 }
\newcommand{\addressofuw}[1]{$^{ff}$ #1 }

\author{
{\large The Super-Kamiokande Collaboration} \\
\bigskip
%
J.~Yoo\authoratseoul,
%
Y.~Ashie\authoraticrr,
S.~Fukuda\authoraticrr,
Y.~Fukuda\authoraticrr,
K.~Ishihara\authoraticrr,
Y.~Itow\authoraticrr,
Y.~Koshio\authoraticrr,
A.~Minamino\authoraticrr,
M.~Miura\authoraticrr,
S.~Moriyama\authoraticrr
% ...
\\
\smallskip
\smallskip
\footnotesize
\it
\addressoficrr{Kamioka Observatory, Institute for Cosmic Ray Research, University of Tokyo, Kamioka, Gifu, 506-1205, Japan}\\
\addressofncen{Research Center for Cosmic Neutrinos, Institute for Cosmic Ray Research, University of Tokyo, Kashiwa, Chiba 277-8582, Japan}\\
\addressofbu{Department of Physics, Boston University, Boston, MA 02215, USA}\\
\addressofbnl{Physics Department, Brookhaven National Laboratory, Upton, NY 11973, USA}\\
\addressofuci{Department of Physics and Astronomy, University of California, Irvine, Irvine, CA 92697-4575, USA }\\
\addressofcsu{Department of Physics, California State University, Dominguez Hills, Carson, CA 90747, USA}\\
\addressofcnu{Department of Physics, Chonnam National University, Kwangju 500-757, Korea}\\
\addressofgmu{Department of Physics, George Mason University, Fairfax, VA 22030, USA }\\
\addressofgifu{Department of Physics, Gifu University, Gifu, Gifu 501-1193, Japan}\\
\addressofuh{Department of Physics and Astronomy, University of Hawaii, Honolulu, HI 96822, USA}\\
\addressofkek{Institute of Particle and Nuclear Studies, High Energy Accelerator Research Organization (KEK), Tsukuba, Ibaraki 305-0801, Japan }\\
\addressofkobe{Department of Physics, Kobe University, Kobe, Hyogo 657-8501, Japan}\\
\addressofkyoto{Department of Physics, Kyoto University, Kyoto 606-8502, Japan}\\
\addressoflanl{Physics Division, P-23, Los Alamos National Laboratory, Los Alamos, NM 87544, USA }\\
\addressoflsu{Department of Physics and Astronomy, Louisiana State University, Baton Rouge, LA 70803, USA }\\
\addressofumd{Department of Physics, University of Maryland, College Park, MD 20742, USA }\\
\addressofmit{Department of Physics, Massachusetts Institute of Technology, Cambridge, MA 02139, USA}\\
\addressofduluth{Department of Physics, University of Minnesota, Duluth, MN 55812-2496, USA}\\
\addressofsuny{Department of Physics and Astronomy, State University of New York, Stony Brook, NY 11794-3800, USA}\\
\addressofnagoya{Department of Physics, Nagoya University, Nagoya, Aichi 464-8602, Japan}\\
\addressofniigata{Department of Physics, Niigata University, Niigata, Niigata 950-2181, Japan }\\
\addressofosaka{Department of Physics, Osaka University, Toyonaka, Osaka 560-0043, Japan}\\
\addressofseoul{Department of Physics, Seoul National University, Seoul 151-742, Korea}\\
\addressofshizuokaseika{Internatinal and Cultural Studies, Shizuoka Seika College, Yaizu, Shizuoka, 425-8611, Japan}\\
\addressofshizuoka{Department of Systems Engineering, Shizuoka University, Hamamatsu, Shizuoka 432-8561, Japan}\\
\addressofskku{Department of Physics, Sungkyunkwan University, Suwon 440-746, Korea}\\
\addressoftohoku{Research Center for Neutrino Science, Tohoku University, Sendai, Miyagi 980-8578, Japan}\\
\addressoftokyo{The University of Tokyo, Tokyo 113-0033, Japan }\\
\addressoftokai{Department of Physics, Tokai University, Hiratsuka, Kanagawa 259-1292, Japan}\\
\addressoftit{Department of Physics, Tokyo Institute for Technology, Meguro, Tokyo 152-8551, Japan }\\
\addressofwarsaw{Institute of Experimental Physics, Warsaw University, 00-681 Warsaw, Poland }\\
\addressofuw{Department of Physics, University of Washington, Seattle, WA 98195-1560, USA}\\
}


\begin{abstract}
	A search for periodic modulations of the solar neutrino flux was performed using the Super-Kamiokande-I data taken from May 31st, 1996 to July 15th, 2001. The detector's capability of measuring the exact time of events, combined with a relatively high yield of solar neutrino events, allows a search for short-time variations in the observed flux. We employed the Lomb test to look for periodic modulations of the observed solar neutrino flux. The obtained periodogram is consistent with statistical fluctuation and no significant periodicity was found.

\end{abstract}
\maketitle

\section{Introduction}
	It is widely accepted that neutrinos have masses based on observation of their oscillation \cite{ref:oscillation} which suggests the presence of physics beyond Standard Model. Neutrino spin orientation would be rotated by the magnetic field in the Sun if neutrinos are Dirac particles with non-vanishing magnetic moments \cite{ref:sfp}. Such neutrino spin-flavor precession will result in a reduction of the observed solar neutrino fluxes because the right-handed neutrinos are sterile and will be unobserved in a detector. On the other hand, if neutrinos are Majorana particles with a flavor-changing transition magnetic moment, the solar magnetic field could flip $\nu_e$ into $\bar{\nu}_\mu$ or $\bar{\nu}_\tau$ and lower the observed solar neutrino fluxes due to the reduced elastic scattering cross section.  However, any such possibility must not allow for a flux of solar $\bar{\nu}_e$ above current experimental limits \cite{ref:sk-nuebar}. Based on the above context, any time variation of the solar magnetic fields could introduce periodic modulations in the observed solar neutrino fluxes. There could be semiannual (seasonal) variations of the observed solar neutrino flux because of the changing magnetic field caused by the 7.25 degree inclination of solar axis with respect to the ecliptic plane. A short-time variation might be expected due to the 27-day rotation of the Sun. \par

	Independent of magnetic effects, the $^{8}$B neutrino flux is sensitive to the core temperature of the Sun, approximately proportional to $T_c^{25}$ \cite{ref:jnb-temp}. If the solar core temperature changes in time, this might induce a time variation in the observed solar neutrino flux.\par
	There has been several attempts to search for possible periodic variations in measured solar neutrino fluxes. The Homestake experiment appeared to exhibit 11-year periodic modulations on the data which anticorrelated with solar activity \cite{ref:homestake}. However, Kamiokande and other experiments have not provided any evidence for a time variation of the neutrino flux outside of statistical fluctuations \cite{ref:kamiokande}. Observation of a periodic modulation in the solar neutrino flux would provide a significant addition to our understanding of solar dynamics and the magnetic properties of neutrinos, and possibly require modification of the Standard Solar Model.\par

	Super-Kamiokande (SK) is a 50 kton water Cherenkov detector located in Kamioka, Japan. Solar neutrino data were collected at SK from May 31st, 1996 to July 15th, 2001 yielding a total detector live time of 1,496 days. This data taking period is known as SK-I. A detailed description of SK can be found elsewhere \cite{ref:solar}\cite{ref:sknim2003}. The solar neutrino signal is extracted from the data using the cos$\theta_{sun}$ distribution, the angular deviation between the Sun and the reconstructed direction of events with total energies ranging between 5 and 20 MeV \cite{ref:sk1258flux}. From the strong forward peak due to elastic scattering of $^{8}$B solar neutrinos on electrons, $22,400 \pm 200$(stat.) solar neutrino interactions were observed in 22.5 ktons of fiducial volume. The relatively high yield of real-time events in SK, 15 events per day, allows a search for short-time periodic modulations in the observed neutrino fluxes. \par

\section{SK 10-day long sampled solar neutrino data}

\begin{figure}[h!]
  \center{\includegraphics[width=15cm]{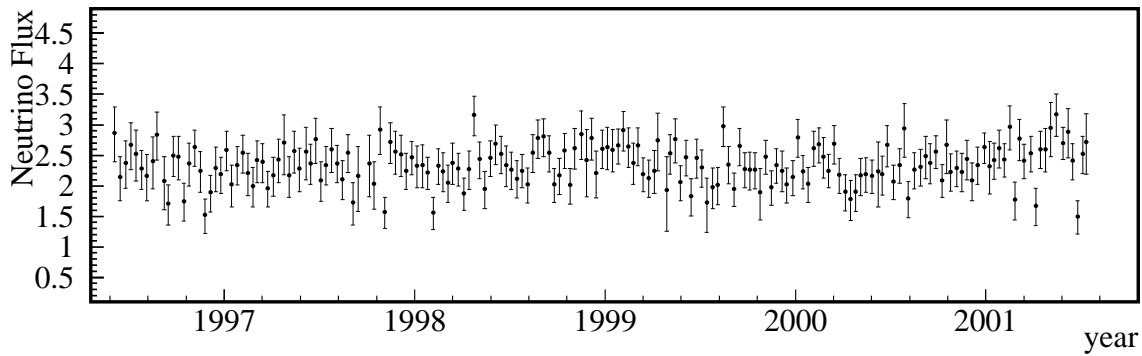}}
  \caption{ \small Measured solar neutrino fluxes of 10-day long samples. The horizontal axis is time (year) from the beginning of the data-taking and the vertical axis is the measured neutrino flux in units of 10$^6$ cm$^{-2}$ s$^{-1}$. The 1/R$^2$ correction is included in the shown neutrino fluxes.}\label{fig:10day}
\end{figure}

	The solar neutrino data, acquired over 1,871 elapsed days from the beginning of data-taking, are divided into roughly 10-day long samples as listed in Table \ref{tab:10day}. The time period of each 10-day sample is chosen from consecutive 10-day periods. Note that the start time of a 10-day sample is determined by the beginning of data-taking on the first day of the sample, and the stop time by the finish of data-taking on the last day. There are on and off periods of data-taking in the 10-day interval and thus the timing of each sample is calculated as a mean of the start and end times and corrected by SK livetime. Hence the mean time is not necessarily an exact division of the time interval for each 10-day sample. Figure \ref{fig:10day} shows the measured solar neutrino fluxes of the 10-day samples. All given uncertainties are statistical and estimated by asymmetric Gaussian approximation of the unbinned maximum likelihood fit to the cos$\theta_{sun}$ distributions. The measured solar neutrino fluxes are corrected for the 1/$R^2$ (squared average distance in units of A.U.) variation caused by the eccentricity of the Earth's orbit around the Sun.\\

\section{Search for Periodicity}

\begin{figure}[t!]
  \center{\includegraphics[width=8cm]{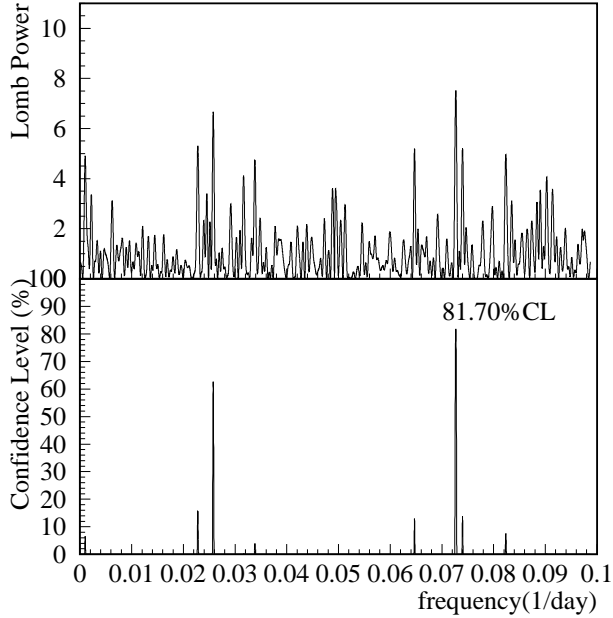}}
  \caption{\label{fig:power} \small A Lomb periodogram of the SK 10-day long solar neutrino data samples. The Lomb power and its corresponding confidence level are given as a function of frequency.}
\end{figure}

\begin{figure}[t!]
  \center{\includegraphics[width=10cm]{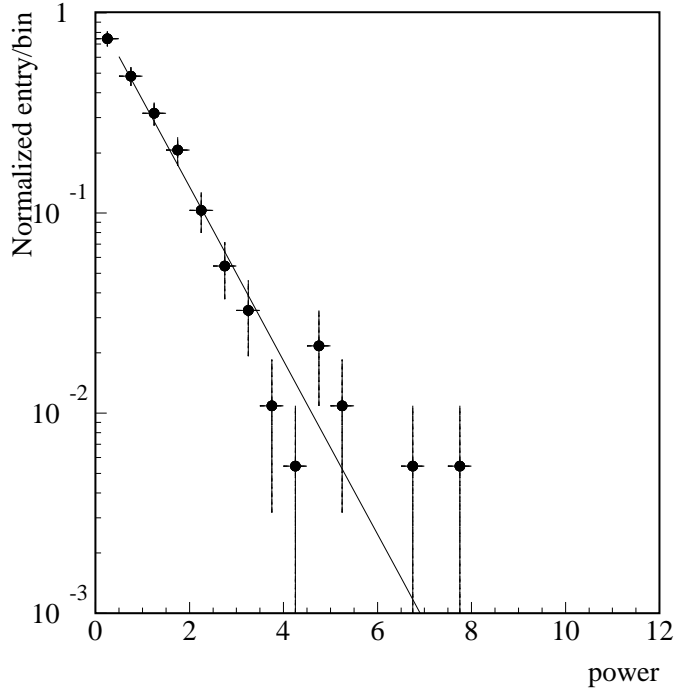}}
  \caption{\label{fig:power-distribution} \small A power distribution of the Lomb periodogram of the SK 10-day long solar neutrino data samples. The solid line shows an expected distribution of no particular periodicity ($e^{-P_N}$).}
\end{figure}

	The Lomb periodogram method \cite{ref:nr13.8}, a spectral analysis for unevenly sampled data, is applied to search for  possible periodicities in the measured 10-day long fluxes. The method finds periodicities based on maximum deviation of data relative to a constant in time. The normalized Lomb power is given by

\begin{equation}\label{eqn:lomb}
P_N (w)\equiv {1 \over {2\sigma^2}} \left\{ {{[\sum_{j}(\phi_j - \bar{\phi})\cos\omega(t_j - \tau)]^2 } \over { \sum_{j}\cos^2\omega(t_j - \tau)}} + {{[\sum_{j}(\phi_j - \bar{\phi})\sin\omega(t_j - \tau)]^2} \over { \sum_{j}\sin^2\omega(t_j - \tau)}}\right\} 
\end{equation}

\vspace{0.5cm}
\noindent
where ${\phi_j}$ is the measured flux in the $j$-th bin, $t_j$ is the mean time with livetime correction in the $j$-th bin, $\omega$ is the frequency being tested, $\bar{\phi}\equiv \sum_{i=1}^N \phi_i / N $, $\sigma^2 \equiv \sum_{i=1}^N (\phi_i - \bar{\phi})^2 / (N-1)$, and offset time $\tau$ is defined by $\tan(2\omega\tau) = {{\sum_j \sin2\omega t_j} / {\sum_j \cos2\omega t_j}}$. Under the null hypothesis, it is expected that the data are Gaussian random values and the Lomb normalized power ($P_N(w)$) is distributed exponentially with unit mean. Here we define the Confidence Level (C.L.) of the given Lomb Power $P_N$ as $(1-e^{-P_N})^M \times 100$\%, the probability of no frequencies being scanned giving values larger than $P_N$ due to random fluctuation. The number of independent frequencies scanned, M, is given by approximately twice of the number of data points, i.e. M = 2 $\times$ 184 = 368. The frequency ranges from 0.00020 day$^{-1}$ to 0.09870 day$^{-1}$ and 18,363 frequencies are scanned in this range. The resulting periodogram is shown in Figure \ref{fig:power} together with its confidence level diagram. The maximum power appears at frequency f=0.0726 day$^{-1}$ (or time period T=13.76 days) with Lomb power 7.51 corresponding to 81.70\% C.L. \par
Figure \ref{fig:power-distribution} shows a normal distribution of the powers in the Lomb periodogram for SK solar neutrino data. Here the powers are calculated for 368 independent frequencies from 0 day$^{-1}$ to 0.1 day$^{-1}$.

\begin{figure}[h!]
  \center{\includegraphics[width=8cm]{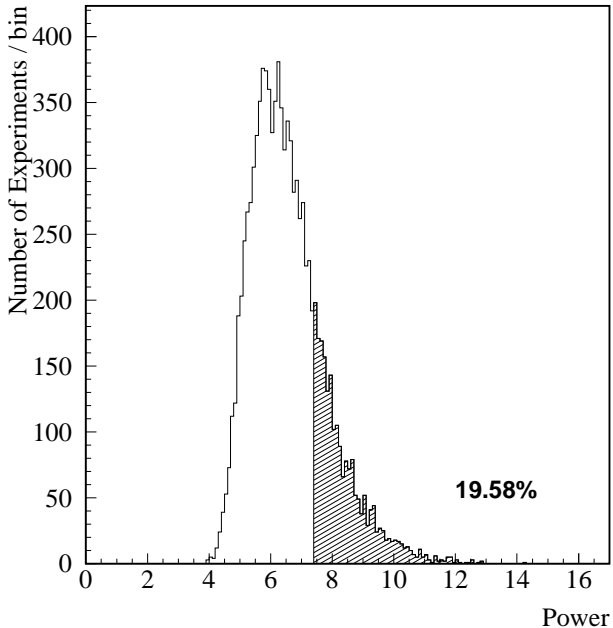}}
  \caption{\label{fig:mcrndm} \small A distribution of maximum powers for 10,000 MC experiment sets. The horizontal axis is Lomb power, and the vertical axis is the number of MC experiment sets.}
\end{figure}

As a consistency check for the confidence level, 10,000 MC experiments are generated based on the observed timing information and the measured solar neutrino fluxes of the 10-day long data samples. The measured solar neutrino flux values are simulated according to a random Gaussian fluctuation. For making these null modulation samples the average measured flux ($2.33 \times 10^6$ cm$^{-2}$ s$^{-1}$) is taken as a Gaussian mean, and the standard deviation of the the measured flux ($0.32 \times 10^6$ cm$^{-2}$ s$^{-1}$) is taken as a Gaussian error \cite{ref:horne_ApJ1986}.
 The Lomb method is applied to each MC experiment to obtain a periodogram. The Lomb power distribution has an exponential distribution as expected for the null modulation. The maximum power of each periodogram is selected. Figure \ref{fig:mcrndm} shows a distribution of maximum powers for the MC experiment sets. Out of 10,000 simulated experiments, 19.58\% have maximum powers larger than 7.51.  This demonstrates that the confidence level for the T=13.76 day period of SK data is consistent with that of no modulation.\par

\vspace{1cm}

\begin{figure}[ht!]
  \center{\includegraphics[width=15cm]{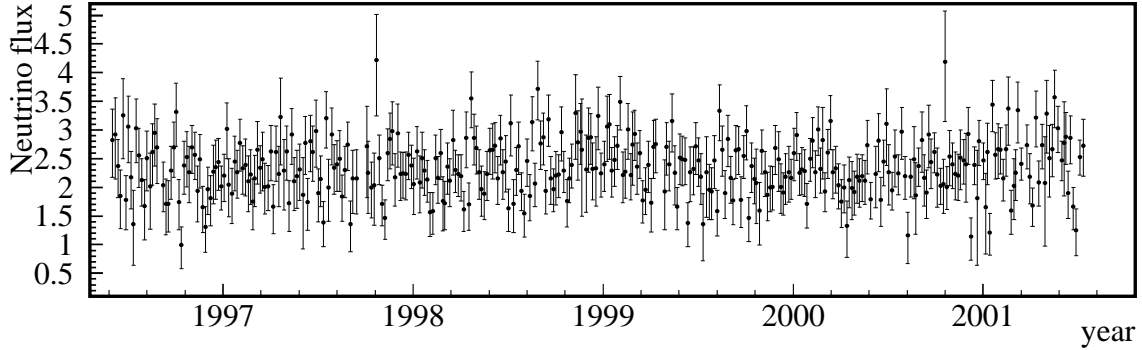}}
  \caption{ \small Measured solar neutrino fluxes of 5-day long samples. The horizontal axis is time (year) from the beginning of the data-taking and the vertical axis is the measured neutrino flux in units of 10$^6$ cm$^{-2}$ s$^{-1}$. The 1/R$^2$ correction is included in the shown neutrino fluxes.}\label{fig:5day}
\end{figure}

\begin{figure}[h!]
  \center{\includegraphics[width=8cm]{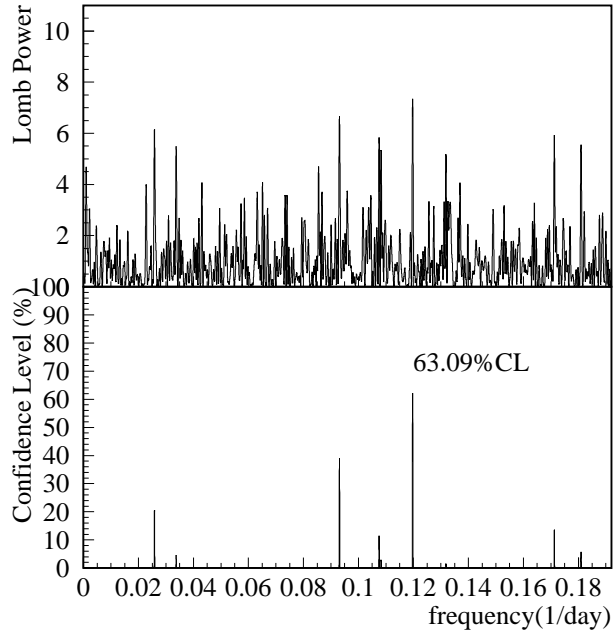}}
  \caption{\label{fig:5day_power} \small A periodogram of the SK 5-day long solar neutrino data samples. The Lomb power and its corresponding confidence level is given as a function of frequency. Note that the searched frequency range is extended up to 0.19 day$^{-1}$. }
\end{figure}

In order to extend the investigation to shorter time modulation more precisely, we repeat the search for periodicities in a time interval of 5 days (see Figure \ref{fig:5day} and Table \ref{tab:5day}) using the Lomb method. Figure \ref{fig:5day_power} shows a periodogram for the 5-day long samples. For the 5-day long sample, the frequency ranges from 0.00020 day$^{-1}$ to 0.19187 day$^{-1}$ and 35,763 frequencies are scanned in this range. The number of independent frequencies scanned is M=716. The maximum Lomb power of 7.35 (63.09\% C.L.), consistent with no modulation, is found at frequency f=0.1197 day$^{-1}$ (T=8.35 days). In this case, the Lomb power for the period of 13.76 days (f=0.0726 day$^{-1}$) is  1.35 and is no longer the maximum. This provides additional confirmation that the 13.76 day period in 10-day long sample is a statistical artifact.

Our results are in clear disagreement with the previous analyses of SK data by Milsztajn \cite{ref:milsztajn2003} and Caldwell and Sturrock \cite{ref:sturrock_ApJ2003}, in which they claim to find a significant periodic variation of our data with 13.76 days' periodicity. The main difference between their analyses and ours is that they have used uniform 10-day binned data in which the time of each data point is exactly centered between the start and stop times of each 10 day period \cite{ref:milsztajn2003} or integrating expected flux from the start time to the end time of each 10-day bin without taking into account breaks of runs \cite{ref:sturrock_ApJ2003}. On the other hand, we use the correct time for the data contained in each 10 day bin based on the livetime weighted mean time as described above. We investigate 5-day binned data as well to find no significant periodic variation.

\section{Sensitivity of Finding a Periodicity vs. Modulation Amplitude}
\begin{figure}[ht!]
  \center{\includegraphics[width=9cm]{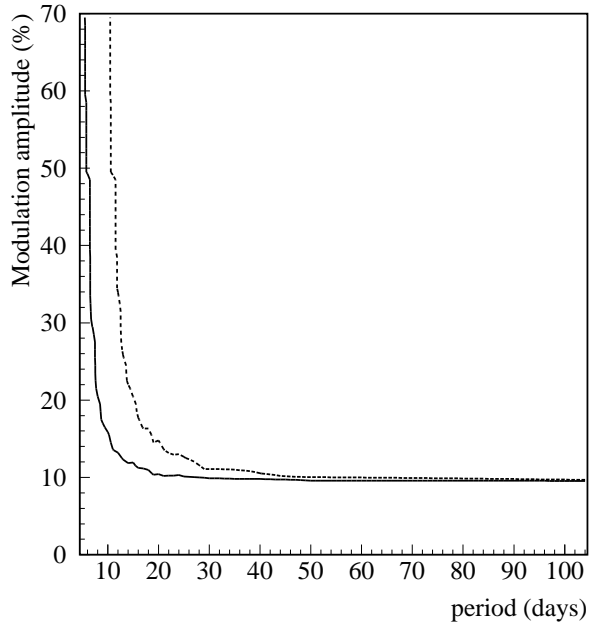}}
  \caption{ \small A 95\% contour plot of the sensitivity to find correct periodicity in 10-day (dotted) and 5-day (solid) long sampled SK solar neutrino data. The horizontal axis is simulated period in days. The vertical axis is modulation amplitude in percentile. Each contour represents 95\% probability of finding the correct periodicity.}\label{fig:sensitivity}
\end{figure}

	We have studied the sensitivity of the SK-I solar neutrino data to find if a true periodicity indeed exists. The sensitivity will depend on the amplitude of modulation because of experimental uncertainties in the SK-I data. One thousand MC experiment sets are generated to simulate the SK 10-day long solar neutrino fluxes for a given period and modulation amplitude. While introducing modulations of periods from 5 days to 500 days and varying the magnitude of modulation from 1\% to 90\% of the averaged total flux ($2.33 \times 10^6$ cm$^{-2}$ s$^{-1}$), we repeatedly generate 1,000 MC experiments at each amplitude. An individual MC experiment set is generated in the following way. The modulated solar neutrino flux is generated for entire SK-1 data-taking period, and the mean flux is calculated for the individual 10-day long time periods. The MC neutrino flux of every 10-day long sample is then generated by a random Gaussian fluctuation from the mean flux value. The Gaussian error is taken as the error in the measured neutrino flux of each 10-day sample as listed in Table \ref{tab:10day}. The Lomb periodogram method is applied to every MC experiment.
  	A reconstructed period is determined by the primary modulation having the maximum Lomb power larger than 9.82 (98\% C.L.). We calculate the probability of correctly finding the generated period at each modulation amplitude. The observed period is assigned to a correct one if its corresponding frequency is consistent with 1/(input period)$\pm$ 0.1/368/2 for the 10-day long sample, to consider the error on the reconstructed period. The probability of reconstructing the input periodicity becomes significantly reduced as the modulation amplitude becomes less than the magnitude of the measured neutrino flux error. The MC study is repeated for 5-day long samples. In the 5-day long sample case, the selection criterion of correct frequency is 1/(input period)$\pm$ 0.2/716/2. Figure \ref{fig:sensitivity} shows 95\% sensitive contours of modulation amplitude vs. period for 10-day and 5-day long samples. The sensitivity is defined as a probability of finding the correct periodicity as a primary modulation.

According to Figure \ref{fig:sensitivity}, the sensitivity of finding the correct period varies rapidly near the sampling time periods of 5 $\sim$ 20 days. In general, shorter modulation periods have worse sensitivity for a given sampling frequency, while the 5-day long sampled data have better sensitivity than those of the 10-day data, especially for shorter modulation periods. Based on the MC study, the Lomb method is expected to find the true periodicity of the SK-I data, if any, with an efficiency of 95\% at a modulation amplitude of 10\% of the averaged flux and a period of longer than 40 days (20 days) in case of the 10-day (5-day) long sample.

\section{Summary}
	We have presented the measured solar neutrino fluxes of 10-day long samples and of 5-day long samples using all 1,496 days of SK-I data. No significant periodicity was found in the SK-I solar neutrino data when a search was made to look for periodic modulations of the observed fluxes using the Lomb method. Based on a MC study, we have obtained the probability of finding a true periodicity in the SK-I data as a function of the modulation magnitude.  The Lomb method should have found a periodic modulation in the SK-I solar neutrino data of 10-day (5-day) long samples if the modulation period were longer than 40 days (20 days) and its magnitude was larger than 10\% of the average measured neutrino fluxes. Based on the observation of no significant periodicity, SK-I data exclude modulations greater than 10\% of the $^8$B neutrino flux arising as a result of more than 0.4\% changes in the solar core temperature, allowing a new measure of the solar core's stability.

\begin{acknowledgments}
The authors acknowledge the cooperation of the Kamioka Mining and 
Smelting Company.
The Super-Kamiokande has been built and operated from funding 
by the Japanese Ministry of Education, Culture, Sports, Science and 
Technology, the U.S. Department of Energy, and 
the U.S. National Science Foundation. 
This work was partially supported by the Korean Research
Foundation (BK21) and the Korea Ministry of Science and
Technology.

\end{acknowledgments}




%
%

\leftmargin 0cm
\rightmargin 0cm
\begin{table}[H]
\caption{\label{tab:10day} SK solar neutrino data divided into 10-day long samples. t$_w$ is mean time with livetime correction. t$_s$ is start time. t$_e$ is end time.  R$^2$ is square of the distance between the Sun and the Earth, in unit of AU$^2$. $\phi_{\nu}$ is measured $^8$B neutrino flux without correction of R$^2$ }
\tiny
\vspace{0cm}
\begin{tabular}{|r|c|c|c|c|c|}
\hline 
No. & t$_w$ & t$_s$ & t$_e$ & R$^2$ & $\phi_{\nu}$ \\
\hline
  1 & 1996/06/04 10:53 & 05/31 04:31 & 06/10 01:19 & 1.029 & 2.79+0.47-0.41 \\
  2 & 1996/06/14 17:56 & 06/10 01:27 & 06/20 03:41 & 1.032 & 2.07+0.37-0.33 \\
  3 & 1996/06/25 13:41 & 06/20 03:45 & 06/30 14:11 & 1.033 & 2.30+0.41-0.36 \\
  4 & 1996/07/05 11:28 & 06/30 14:16 & 07/10 03:18 & 1.034 & 2.58+0.40-0.35 \\
  5 & 1996/07/15 14:18 & 07/10 03:30 & 07/20 01:19 & 1.033 & 2.44+0.43-0.38 \\
  6 & 1996/07/25 18:06 & 07/20 01:28 & 07/31 08:19 & 1.032 & 2.21+0.33-0.29 \\
  7 & 1996/08/05 10:46 & 07/31 16:06 & 08/10 00:35 & 1.029 & 2.10+0.39-0.34 \\
  8 & 1996/08/16 12:09 & 08/10 20:31 & 08/20 06:16 & 1.025 & 2.34+0.44-0.39 \\
  9 & 1996/08/24 14:31 & 08/20 06:33 & 08/31 11:47 & 1.022 & 2.78+0.41-0.36 \\
 10 & 1996/09/07 21:16 & 08/31 11:51 & 09/10 11:28 & 1.015 & 2.05+0.46-0.39 \\
 11 & 1996/09/15 11:16 & 09/10 11:53 & 09/20 00:38 & 1.011 & 1.69+0.35-0.30 \\
 12 & 1996/09/25 10:01 & 09/20 05:00 & 09/30 11:13 & 1.006 & 2.49+0.35-0.31 \\
 13 & 1996/10/05 04:43 & 09/30 11:14 & 10/10 10:31 & 1.000 & 2.48+0.38-0.33 \\
 14 & 1996/10/15 11:21 & 10/10 10:42 & 10/20 16:22 & 0.994 & 1.76+0.33-0.29 \\
 15 & 1996/10/25 05:19 & 10/20 17:31 & 10/30 03:19 & 0.989 & 2.40+0.37-0.33 \\
 16 & 1996/11/05 01:54 & 10/30 03:47 & 11/10 22:14 & 0.983 & 2.68+0.32-0.28 \\
 17 & 1996/11/15 19:54 & 11/10 22:17 & 11/20 11:37 & 0.978 & 2.30+0.37-0.32 \\
 18 & 1996/11/25 08:59 & 11/20 12:34 & 11/30 01:02 & 0.974 & 1.56+0.31-0.27 \\
 19 & 1996/12/05 08:16 & 11/30 01:07 & 12/10 07:00 & 0.971 & 1.95+0.33-0.29 \\
 20 & 1996/12/15 20:58 & 12/10 07:04 & 12/20 16:53 & 0.969 & 2.36+0.40-0.35 \\
 21 & 1996/12/25 22:35 & 12/20 17:14 & 12/31 04:42 & 0.967 & 2.26+0.32-0.29 \\
 22 & 1997/01/05 15:32 & 12/31 04:46 & 01/10 14:11 & 0.967 & 2.68+0.35-0.31 \\
 23 & 1997/01/15 12:41 & 01/10 14:15 & 01/20 09:04 & 0.968 & 2.09+0.38-0.33 \\
 24 & 1997/01/26 03:51 & 01/20 16:50 & 01/31 08:09 & 0.969 & 2.41+0.34-0.31 \\
 25 & 1997/02/05 12:25 & 01/31 08:58 & 02/10 16:01 & 0.972 & 2.61+0.34-0.30 \\
 26 & 1997/02/15 09:40 & 02/10 16:04 & 02/20 06:56 & 0.976 & 2.26+0.38-0.33 \\
 27 & 1997/02/24 17:56 & 02/20 08:22 & 02/28 22:36 & 0.980 & 2.04+0.36-0.31 \\
 28 & 1997/03/05 07:33 & 02/28 22:39 & 03/10 09:39 & 0.984 & 2.46+0.38-0.33 \\
 29 & 1997/03/15 04:28 & 03/10 09:46 & 03/20 02:16 & 0.989 & 2.42+0.34-0.30 \\
 30 & 1997/03/25 16:07 & 03/20 02:26 & 03/31 09:21 & 0.995 & 1.97+0.31-0.28 \\
 31 & 1997/04/05 10:10 & 03/31 21:11 & 04/10 03:03 & 1.001 & 2.17+0.34-0.30 \\
 32 & 1997/04/15 06:10 & 04/10 03:46 & 04/20 04:41 & 1.006 & 2.42+0.38-0.33 \\
 33 & 1997/04/25 08:16 & 04/20 07:17 & 04/30 00:22 & 1.012 & 2.68+0.52-0.44 \\
 34 & 1997/05/05 07:41 & 04/30 00:31 & 05/10 03:37 & 1.017 & 2.13+0.35-0.30 \\
 35 & 1997/05/15 01:11 & 05/10 04:19 & 05/20 00:38 & 1.022 & 2.51+0.35-0.31 \\
 36 & 1997/05/25 08:20 & 05/20 00:41 & 05/31 05:47 & 1.026 & 2.22+0.37-0.33 \\
 37 & 1997/06/06 17:41 & 05/31 06:10 & 06/10 02:33 & 1.030 & 2.49+0.45-0.39 \\
 38 & 1997/06/15 05:51 & 06/10 03:33 & 06/20 12:17 & 1.032 & 2.30+0.34-0.30 \\
 39 & 1997/06/25 02:15 & 06/20 12:22 & 06/30 02:28 & 1.033 & 2.67+0.38-0.34 \\
 40 & 1997/07/05 12:15 & 06/30 02:31 & 07/10 07:24 & 1.034 & 2.02+0.33-0.29 \\
 41 & 1997/07/15 07:04 & 07/10 07:29 & 07/20 12:47 & 1.033 & 2.27+0.32-0.29 \\
 42 & 1997/07/25 19:34 & 07/20 12:55 & 07/31 00:40 & 1.032 & 2.52+0.37-0.33 \\
 43 & 1997/08/05 04:59 & 07/31 01:04 & 08/10 05:33 & 1.029 & 2.30+0.31-0.28 \\
 44 & 1997/08/15 16:59 & 08/10 05:38 & 08/20 14:35 & 1.026 & 2.06+0.33-0.29 \\
 45 & 1997/08/26 02:23 & 08/20 14:45 & 08/31 13:09 & 1.021 & 2.49+0.31-0.29 \\
 46 & 1997/09/04 16:06 & 08/31 13:11 & 09/10 04:48 & 1.017 & 1.70+0.36-0.31 \\
 47 & 1997/09/14 11:21 & 09/10 04:50 & 09/20 01:01 & 1.012 & 2.13+0.57-0.48 \\
 48 & 1997/10/05 21:24 & 09/30 00:42 & 10/10 18:48 & 1.000 & 2.37+0.54-0.46 \\
 49 & 1997/10/15 10:10 & 10/10 18:50 & 10/20 16:16 & 0.994 & 2.04+0.41-0.36 \\
 50 & 1997/10/26 19:13 & 10/20 18:13 & 10/30 14:40 & 0.988 & 2.96+0.42-0.37 \\
 51 & 1997/11/05 02:59 & 10/30 14:49 & 11/10 18:55 & 0.983 & 1.60+0.27-0.24 \\
 52 & 1997/11/15 15:50 & 11/11 00:53 & 11/20 19:59 & 0.978 & 2.78+0.36-0.32 \\
 53 & 1997/11/25 00:37 & 11/20 20:02 & 11/30 02:45 & 0.975 & 2.63+0.38-0.34 \\
 54 & 1997/12/05 15:43 & 11/30 04:04 & 12/10 11:35 & 0.971 & 2.58+0.37-0.33 \\
 55 & 1997/12/15 08:05 & 12/10 11:38 & 12/20 13:00 & 0.969 & 2.32+0.32-0.29 \\
 56 & 1997/12/26 04:25 & 12/20 13:14 & 12/31 21:10 & 0.967 & 2.55+0.30-0.27 \\
 57 & 1998/01/05 20:18 & 12/31 21:12 & 01/10 23:32 & 0.967 & 2.41+0.37-0.33 \\
 58 & 1998/01/16 01:49 & 01/10 23:34 & 01/20 08:45 & 0.968 & 2.42+0.38-0.34 \\
 59 & 1998/01/25 21:40 & 01/20 09:11 & 01/31 11:16 & 0.969 & 2.29+0.29-0.26 \\
 60 & 1998/02/05 10:24 & 01/31 11:18 & 02/10 06:59 & 0.972 & 1.61+0.29-0.26 \\
\hline
\end{tabular}
\begin{tabular}{|r|c|c|c|c|c|}
\hline
No. & t$_w$ & t$_s$ & t$_e$ & $R^2$ & $\phi_{\nu}$ \\
\hline
 61 & 1998/02/15 07:17 & 02/10 07:02 & 02/20 08:10 & 0.976 & 2.39+0.32-0.29 \\
 62 & 1998/02/24 20:51 & 02/20 09:45 & 02/28 17:32 & 0.980 & 2.29+0.35-0.31 \\
 63 & 1998/03/05 13:24 & 02/28 17:35 & 03/10 06:37 & 0.984 & 2.08+0.33-0.29 \\
 64 & 1998/03/14 21:02 & 03/10 06:41 & 03/20 00:39 & 0.989 & 2.41+0.36-0.32 \\
 65 & 1998/03/25 20:16 & 03/20 00:42 & 03/31 15:49 & 0.995 & 2.29+0.28-0.25 \\
 66 & 1998/04/05 20:52 & 03/31 15:51 & 04/10 22:50 & 1.001 & 1.88+0.28-0.25 \\
 67 & 1998/04/15 14:59 & 04/10 22:52 & 04/20 10:41 & 1.007 & 2.25+0.33-0.30 \\
 68 & 1998/04/25 09:40 & 04/20 11:35 & 04/30 10:27 & 1.012 & 3.12+0.34-0.31 \\
 69 & 1998/05/05 07:58 & 04/30 10:48 & 05/10 06:12 & 1.017 & 2.40+0.31-0.28 \\
 70 & 1998/05/15 12:26 & 05/10 06:14 & 05/20 12:03 & 1.022 & 1.91+0.31-0.28 \\
 71 & 1998/05/26 04:27 & 05/20 12:06 & 05/31 16:14 & 1.026 & 2.40+0.30-0.27 \\
 72 & 1998/06/05 12:29 & 05/31 16:15 & 06/10 07:54 & 1.029 & 2.61+0.33-0.30 \\
 73 & 1998/06/15 09:00 & 06/10 07:57 & 06/20 11:44 & 1.032 & 2.44+0.31-0.28 \\
 74 & 1998/06/25 04:52 & 06/20 11:48 & 06/30 00:00 & 1.033 & 2.27+0.34-0.30 \\
 75 & 1998/07/05 04:02 & 06/30 00:02 & 07/10 23:08 & 1.034 & 2.19+0.31-0.28 \\
 76 & 1998/07/15 20:04 & 07/10 23:13 & 07/20 13:25 & 1.033 & 2.05+0.31-0.28 \\
 77 & 1998/07/26 03:52 & 07/20 13:26 & 07/31 19:08 & 1.032 & 2.18+0.27-0.25 \\
 78 & 1998/08/05 15:29 & 07/31 19:13 & 08/10 13:48 & 1.029 & 1.96+0.29-0.26 \\
 79 & 1998/08/15 16:56 & 08/10 13:51 & 08/20 13:30 & 1.026 & 2.48+0.31-0.29 \\
 80 & 1998/08/25 11:47 & 08/20 13:32 & 08/31 08:08 & 1.022 & 2.72+0.32-0.29 \\
 81 & 1998/09/05 10:26 & 08/31 08:29 & 09/10 15:07 & 1.017 & 2.76+0.32-0.29 \\
 82 & 1998/09/15 17:10 & 09/10 15:17 & 09/20 15:19 & 1.011 & 2.51+0.31-0.28 \\
 83 & 1998/09/25 10:04 & 09/20 15:20 & 09/30 05:05 & 1.006 & 2.01+0.29-0.26 \\
 84 & 1998/10/05 16:02 & 09/30 05:10 & 10/10 15:33 & 1.000 & 2.17+0.31-0.28 \\
 85 & 1998/10/15 18:00 & 10/10 15:35 & 10/20 20:36 & 0.994 & 2.60+0.30-0.28 \\
 86 & 1998/10/26 02:24 & 10/20 20:39 & 10/30 07:51 & 0.989 & 2.03+0.31-0.27 \\
 87 & 1998/11/03 23:30 & 10/30 07:54 & 11/10 15:13 & 0.984 & 2.66+0.36-0.33 \\
 88 & 1998/11/16 11:43 & 11/10 18:00 & 11/20 13:50 & 0.978 & 2.91+0.43-0.39 \\
 89 & 1998/11/26 23:12 & 11/20 13:52 & 11/30 08:38 & 0.974 & 2.48+0.62-0.51 \\
 90 & 1998/12/06 12:54 & 11/30 09:28 & 12/10 12:26 & 0.971 & 2.86+0.37-0.34 \\
 91 & 1998/12/14 19:58 & 12/10 12:31 & 12/20 05:27 & 0.969 & 2.28+0.42-0.37 \\
 92 & 1998/12/26 10:07 & 12/20 05:31 & 12/31 13:01 & 0.967 & 2.70+0.34-0.31 \\
 93 & 1999/01/05 10:19 & 12/31 13:58 & 01/10 20:01 & 0.967 & 2.72+0.38-0.34 \\
 94 & 1999/01/15 19:32 & 01/10 20:03 & 01/20 00:03 & 0.968 & 2.68+0.37-0.34 \\
 95 & 1999/01/26 06:10 & 01/20 00:45 & 01/31 18:46 & 0.969 & 2.74+0.31-0.28 \\
 96 & 1999/02/05 07:05 & 01/31 18:47 & 02/10 09:15 & 0.972 & 2.99+0.35-0.32 \\
 97 & 1999/02/15 06:16 & 02/10 16:30 & 02/20 00:00 & 0.975 & 2.71+0.35-0.32 \\
 98 & 1999/02/24 03:17 & 02/20 00:03 & 02/28 02:13 & 0.979 & 2.43+0.37-0.33 \\
 99 & 1999/03/04 17:06 & 02/28 02:16 & 03/10 00:33 & 0.983 & 2.71+0.34-0.30 \\
100 & 1999/03/15 04:46 & 03/10 01:03 & 03/20 08:31 & 0.989 & 2.21+0.29-0.27 \\
101 & 1999/03/25 23:27 & 03/20 08:34 & 03/31 16:11 & 0.995 & 2.14+0.32-0.29 \\
102 & 1999/04/05 15:50 & 04/01 01:42 & 04/10 00:02 & 1.001 & 2.25+0.37-0.33 \\
103 & 1999/04/12 05:55 & 04/10 00:22 & 04/20 00:43 & 1.004 & 2.74+0.50-0.44 \\
104 & 1999/04/29 15:01 & 04/20 00:49 & 04/30 19:59 & 1.014 & 1.90+0.66-0.54 \\
105 & 1999/05/05 10:49 & 04/30 20:00 & 05/10 04:14 & 1.017 & 2.49+0.33-0.30 \\
106 & 1999/05/15 18:36 & 05/10 04:15 & 05/20 16:02 & 1.022 & 2.70+0.37-0.33 \\
107 & 1999/05/25 16:49 & 05/20 16:06 & 05/31 04:51 & 1.026 & 2.01+0.29-0.26 \\
108 & 1999/06/05 12:48 & 05/31 04:53 & 06/10 18:57 & 1.029 & 2.40+0.30-0.27 \\
109 & 1999/06/15 02:11 & 06/10 18:59 & 06/20 02:39 & 1.032 & 1.77+0.31-0.28 \\
110 & 1999/06/25 11:15 & 06/20 02:43 & 06/30 05:21 & 1.033 & 2.38+0.32-0.29 \\
111 & 1999/07/05 07:36 & 06/30 05:39 & 07/10 08:12 & 1.034 & 2.22+0.31-0.28 \\
112 & 1999/07/14 23:27 & 07/10 08:18 & 07/20 00:09 & 1.033 & 1.67+0.47-0.39 \\
113 & 1999/07/26 11:28 & 07/20 00:11 & 07/31 02:30 & 1.032 & 1.92+0.34-0.30 \\
114 & 1999/08/05 11:20 & 07/31 02:39 & 08/10 15:46 & 1.029 & 1.95+0.31-0.28 \\
115 & 1999/08/15 16:24 & 08/10 15:48 & 08/20 17:18 & 1.026 & 2.90+0.33-0.30 \\
116 & 1999/08/25 22:26 & 08/20 17:19 & 08/31 11:54 & 1.022 & 2.30+0.31-0.28 \\
117 & 1999/09/05 11:36 & 08/31 11:57 & 09/10 13:27 & 1.017 & 1.92+0.29-0.26 \\
118 & 1999/09/15 19:28 & 09/10 13:28 & 09/20 21:40 & 1.011 & 2.62+0.32-0.29 \\
119 & 1999/09/25 21:49 & 09/20 21:41 & 09/30 20:40 & 1.006 & 2.26+0.30-0.27 \\
120 & 1999/10/05 06:58 & 09/30 20:42 & 10/10 12:04 & 1.000 & 2.26+0.33-0.29 \\
\hline
\end{tabular}
\end{table}
\begin{table}[h]
\tiny
\begin{tabular}{|r|c|c|c|c|c|}
\hline
No. & t$_w$ & t$_s$ & t$_e$ & $R^2$ & $\phi_{\nu}$ \\
\hline
121 & 1999/10/15 12:46 & 10/10 12:09 & 10/21 00:12 & 0.995 & 2.27+0.31-0.28 \\
122 & 1999/10/25 11:57 & 10/21 00:14 & 10/30 10:26 & 0.989 & 1.91+0.45-0.39 \\
123 & 1999/11/05 00:08 & 10/30 10:28 & 11/10 14:56 & 0.984 & 2.52+0.29-0.27 \\
124 & 1999/11/15 17:32 & 11/10 14:57 & 11/20 09:14 & 0.979 & 2.02+0.31-0.28 \\
125 & 1999/11/25 06:07 & 11/20 09:15 & 11/30 12:44 & 0.975 & 2.40+0.31-0.28 \\
126 & 1999/12/06 12:36 & 11/30 18:24 & 12/10 03:53 & 0.971 & 2.32+0.36-0.33 \\
127 & 1999/12/15 04:00 & 12/10 08:07 & 12/20 09:21 & 0.969 & 2.08+0.31-0.28 \\
128 & 1999/12/27 02:01 & 12/20 10:20 & 12/31 20:06 & 0.967 & 2.21+0.31-0.28 \\
129 & 2000/01/05 10:22 & 12/31 20:07 & 01/10 01:56 & 0.967 & 2.89+0.33-0.30 \\
130 & 2000/01/15 05:01 & 01/10 01:58 & 01/20 14:16 & 0.967 & 2.32+0.30-0.27 \\
131 & 2000/01/25 07:50 & 01/20 14:33 & 01/31 01:28 & 0.969 & 2.09+0.31-0.28 \\
132 & 2000/02/05 06:15 & 01/31 01:39 & 02/10 08:41 & 0.972 & 2.70+0.31-0.28 \\
133 & 2000/02/15 10:41 & 02/10 08:42 & 02/20 10:48 & 0.975 & 2.75+0.31-0.28 \\
134 & 2000/02/24 05:01 & 02/20 10:49 & 02/28 01:22 & 0.979 & 2.53+0.36-0.32 \\
135 & 2000/03/04 19:09 & 02/28 01:23 & 03/10 08:51 & 0.984 & 2.29+0.28-0.26 \\
136 & 2000/03/15 07:08 & 03/10 08:55 & 03/20 02:14 & 0.989 & 2.72+0.33-0.30 \\
137 & 2000/03/25 17:49 & 03/20 02:23 & 03/31 09:33 & 0.995 & 2.19+0.30-0.27 \\
138 & 2000/04/05 21:39 & 03/31 20:08 & 04/10 14:37 & 1.001 & 1.90+0.31-0.28 \\
139 & 2000/04/16 05:33 & 04/10 15:24 & 04/20 08:57 & 1.007 & 1.77+0.35-0.31 \\
140 & 2000/04/26 03:26 & 04/20 14:20 & 04/30 17:04 & 1.013 & 1.88+0.32-0.28 \\
141 & 2000/05/05 15:10 & 04/30 17:20 & 05/10 08:26 & 1.017 & 2.13+0.32-0.28 \\
142 & 2000/05/15 17:26 & 05/10 08:28 & 05/20 16:32 & 1.022 & 2.14+0.29-0.26 \\
143 & 2000/05/27 02:16 & 05/20 16:33 & 06/02 15:13 & 1.027 & 2.10+0.27-0.25 \\
144 & 2000/06/08 00:31 & 06/02 15:14 & 06/11 13:08 & 1.030 & 2.17+0.57-0.47 \\
145 & 2000/06/16 00:50 & 06/11 13:09 & 06/20 08:52 & 1.032 & 2.12+0.33-0.29 \\
146 & 2000/06/25 05:19 & 06/20 08:56 & 06/30 17:36 & 1.033 & 2.58+0.35-0.31 \\
147 & 2000/07/07 14:22 & 07/01 17:50 & 07/13 14:47 & 1.034 & 2.00+0.29-0.26 \\
148 & 2000/07/19 08:53 & 07/13 14:49 & 07/25 08:09 & 1.033 & 2.27+0.29-0.26 \\
149 & 2000/07/28 11:07 & 07/25 12:46 & 07/31 01:08 & 1.031 & 2.85+0.46-0.40 \\
150 & 2000/08/05 05:32 & 07/31 01:10 & 08/10 10:40 & 1.029 & 1.74+0.30-0.27 \\
151 & 2000/08/16 10:22 & 08/10 10:47 & 08/22 08:03 & 1.025 & 2.20+0.30-0.27 \\
152 & 2000/08/28 15:52 & 08/22 08:11 & 09/03 03:28 & 1.020 & 2.26+0.30-0.28 \\
153 & 2000/09/07 05:54 & 09/03 03:30 & 09/11 09:38 & 1.016 & 2.45+0.36-0.32 \\
154 & 2000/09/15 21:46 & 09/11 15:22 & 09/20 04:05 & 1.011 & 2.35+0.35-0.31 \\
155 & 2000/09/26 20:26 & 09/20 04:09 & 10/03 03:01 & 1.005 & 2.55+0.28-0.26 \\
156 & 2000/10/09 06:41 & 10/03 03:02 & 10/15 02:45 & 0.998 & 2.09+0.28-0.26 \\
157 & 2000/10/17 12:17 & 10/15 02:49 & 10/20 01:19 & 0.993 & 2.69+0.47-0.41 \\
158 & 2000/10/25 13:44 & 10/20 01:22 & 10/31 07:33 & 0.989 & 2.25+0.32-0.29 \\
159 & 2000/11/05 21:59 & 10/31 07:34 & 11/11 06:38 & 0.983 & 2.33+0.31-0.28 \\
160 & 2000/11/15 20:05 & 11/11 07:11 & 11/21 15:44 & 0.978 & 2.28+0.34-0.31 \\
161 & 2000/11/25 20:58 & 11/21 17:26 & 11/30 08:10 & 0.974 & 2.51+0.35-0.32 \\
162 & 2000/12/05 16:39 & 11/30 08:26 & 12/10 02:44 & 0.971 & 2.15+0.34-0.30 \\
163 & 2000/12/15 22:02 & 12/10 02:46 & 12/20 23:17 & 0.969 & 2.42+0.33-0.30 \\
164 & 2000/12/28 23:32 & 12/20 23:32 & 01/05 09:49 & 0.967 & 2.72+0.27-0.25 \\
165 & 2001/01/08 14:58 & 01/05 09:52 & 01/11 16:44 & 0.967 & 2.40+0.47-0.42 \\
166 & 2001/01/16 17:27 & 01/11 16:46 & 01/22 03:19 & 0.968 & 2.50+0.32-0.29 \\
167 & 2001/01/26 16:08 & 01/22 03:21 & 01/31 03:25 & 0.970 & 2.70+0.34-0.30 \\
168 & 2001/02/05 15:56 & 01/31 03:28 & 02/11 06:17 & 0.972 & 2.50+0.30-0.27 \\
169 & 2001/02/15 17:47 & 02/11 06:19 & 02/20 01:26 & 0.976 & 3.04+0.39-0.35 \\
170 & 2001/02/25 12:14 & 02/20 01:35 & 03/02 08:36 & 0.980 & 1.81+0.34-0.30 \\
171 & 2001/03/06 06:04 & 03/02 08:38 & 03/10 04:39 & 0.984 & 2.81+0.37-0.33 \\
172 & 2001/03/15 08:39 & 03/10 04:44 & 03/23 09:21 & 0.989 & 2.44+0.29-0.27 \\
173 & 2001/03/28 15:55 & 03/23 09:24 & 04/03 00:02 & 0.996 & 2.54+0.30-0.28 \\
174 & 2001/04/06 18:16 & 04/03 00:05 & 04/10 14:47 & 1.002 & 1.67+0.32-0.29 \\
175 & 2001/04/15 19:22 & 04/10 14:50 & 04/20 22:30 & 1.007 & 2.58+0.32-0.29 \\
176 & 2001/04/25 04:17 & 04/20 22:33 & 04/30 09:36 & 1.012 & 2.57+0.37-0.34 \\
177 & 2001/05/05 15:58 & 04/30 22:18 & 05/10 16:53 & 1.017 & 2.90+0.46-0.41 \\
178 & 2001/05/16 06:07 & 05/10 16:55 & 05/21 01:50 & 1.022 & 3.10+0.35-0.32 \\
179 & 2001/05/28 20:40 & 05/21 01:52 & 06/04 20:38 & 1.027 & 2.63+0.26-0.25 \\
180 & 2001/06/07 23:38 & 06/04 20:41 & 06/11 02:23 & 1.030 & 2.80+0.41-0.37 \\
181 & 2001/06/15 23:08 & 06/11 02:26 & 06/20 16:35 & 1.032 & 2.34+0.30-0.27 \\
182 & 2001/06/25 17:45 & 06/20 16:38 & 06/30 20:07 & 1.033 & 1.45+0.28-0.25 \\
183 & 2001/07/05 14:02 & 06/30 20:08 & 07/10 11:38 & 1.034 & 2.45+0.31-0.28 \\
184 & 2001/07/12 16:22 & 07/10 12:08 & 07/15 08:34 & 1.033 & 2.63+0.51-0.45 \\

\hline
\end{tabular}
\end{table}
\rightmargin 0cm

%
%

\leftmargin 0cm
\rightmargin 0cm
\begin{table}[h!]
\caption{\label{tab:5day} SK solar neutrino data divided into 5-day long samples. t$_w$ is mean time with livetime correction. t$_s$ is start time. t$_e$ is end time.  R$^2$ is square of the distance between the Sun and the Earth, in unit of AU$^2$. $\phi_{\nu}$ is measured $^8$B neutrino flux without correction of R$^2$ }
\tiny
\vspace{0cm}
\begin{tabular}{|r|c|c|c|c|c|}
\hline 
No. & t$_w$ & t$_s$ & t$_e$ & R$^2$ & $\phi_{\nu}$ \\
\hline
   1 & 1996/06/02 03:46 & 05/31 04:31 & 06/05 08:49 & 1.029 & 2.74+0.63-0.53 \\
   2 & 1996/06/07 14:58 & 06/05 14:13 & 06/10 01:19 & 1.030 & 2.83+0.75-0.62 \\
   3 & 1996/06/12 12:40 & 06/10 01:27 & 06/15 04:36 & 1.031 & 2.30+0.53-0.45 \\
   4 & 1996/06/17 16:37 & 06/15 09:14 & 06/20 03:41 & 1.032 & 1.79+0.55-0.44 \\
   5 & 1996/06/22 03:07 & 06/20 03:45 & 06/25 04:55 & 1.033 & 3.15+0.74-0.61 \\
   6 & 1996/06/27 23:30 & 06/25 05:02 & 06/30 14:11 & 1.033 & 1.72+0.50-0.40 \\
   7 & 1996/07/02 20:29 & 06/30 14:16 & 07/05 05:29 & 1.034 & 2.96+0.61-0.51 \\
   8 & 1996/07/07 18:45 & 07/05 05:36 & 07/10 03:18 & 1.034 & 2.10+0.53-0.44 \\
   9 & 1996/07/12 11:30 & 07/10 03:30 & 07/15 00:22 & 1.033 & 1.31+0.70-0.57 \\
  10 & 1996/07/17 22:07 & 07/15 00:25 & 07/20 01:19 & 1.033 & 2.94+0.58-0.50 \\
  11 & 1996/07/22 11:35 & 07/20 01:28 & 07/25 00:34 & 1.032 & 2.48+0.54-0.45 \\
  12 & 1996/07/28 04:01 & 07/25 00:59 & 07/31 08:19 & 1.031 & 2.05+0.43-0.37 \\
  13 & 1996/08/03 02:39 & 07/31 16:06 & 08/05 03:46 & 1.030 & 1.62+0.58-0.45 \\
  14 & 1996/08/07 07:33 & 08/05 03:50 & 08/10 00:35 & 1.028 & 2.45+0.55-0.46 \\
  15 & 1996/08/13 21:16 & 08/10 20:31 & 08/15 03:23 & 1.026 & 1.96+0.73-0.57 \\
  16 & 1996/08/17 20:59 & 08/15 03:58 & 08/20 06:16 & 1.025 & 2.55+0.57-0.48 \\
  17 & 1996/08/22 14:58 & 08/20 06:33 & 08/25 08:37 & 1.023 & 2.89+0.57-0.49 \\
  18 & 1996/08/27 03:25 & 08/25 08:56 & 08/31 11:47 & 1.021 & 2.63+0.61-0.50 \\
  19 & 1996/09/08 00:02 & 09/05 02:47 & 09/10 11:28 & 1.015 & 2.01+0.46-0.39 \\
  20 & 1996/09/12 19:05 & 09/10 11:53 & 09/15 02:08 & 1.013 & 1.69+0.53-0.43 \\
  21 & 1996/09/17 15:43 & 09/15 03:37 & 09/20 00:38 & 1.010 & 1.70+0.49-0.40 \\
  22 & 1996/09/22 18:24 & 09/20 05:00 & 09/25 03:55 & 1.007 & 2.28+0.49-0.42 \\
  23 & 1996/09/28 00:31 & 09/25 04:27 & 09/30 11:13 & 1.004 & 2.68+0.52-0.45 \\
  24 & 1996/10/02 11:45 & 09/30 11:14 & 10/05 01:13 & 1.002 & 3.31+0.60-0.51 \\
  25 & 1996/10/07 18:36 & 10/05 01:46 & 10/10 10:31 & 0.999 & 1.75+0.48-0.39 \\
  26 & 1996/10/12 20:47 & 10/10 10:42 & 10/15 08:16 & 0.996 & 1.00+0.41-0.33 \\
  27 & 1996/10/18 01:24 & 10/15 09:48 & 10/20 16:22 & 0.993 & 2.39+0.50-0.42 \\
  28 & 1996/10/23 02:45 & 10/20 17:31 & 10/25 12:04 & 0.990 & 2.55+0.52-0.44 \\
  29 & 1996/10/27 18:29 & 10/25 13:08 & 10/30 03:19 & 0.987 & 2.28+0.54-0.45 \\
  30 & 1996/11/02 02:38 & 10/30 03:47 & 11/05 00:53 & 0.985 & 2.73+0.46-0.39 \\
  31 & 1996/11/08 00:22 & 11/05 01:30 & 11/10 22:14 & 0.982 & 2.62+0.45-0.39 \\
  32 & 1996/11/12 18:47 & 11/10 22:17 & 11/15 06:11 & 0.979 & 1.98+0.55-0.44 \\
  33 & 1996/11/17 21:41 & 11/15 06:45 & 11/20 11:37 & 0.977 & 2.55+0.52-0.44 \\
  34 & 1996/11/23 04:20 & 11/20 12:34 & 11/25 16:01 & 0.975 & 1.69+0.46-0.37 \\
  35 & 1996/11/27 19:50 & 11/25 16:05 & 11/30 01:02 & 0.973 & 1.35+0.45-0.36 \\
  36 & 1996/12/02 20:53 & 11/30 01:07 & 12/05 18:46 & 0.972 & 2.02+0.47-0.39 \\
  37 & 1996/12/08 04:11 & 12/05 18:50 & 12/10 07:00 & 0.970 & 1.87+0.49-0.40 \\
  38 & 1996/12/13 07:39 & 12/10 07:04 & 12/15 22:32 & 0.969 & 2.35+0.58-0.48 \\
  39 & 1996/12/18 04:07 & 12/15 22:37 & 12/20 16:53 & 0.968 & 2.43+0.58-0.49 \\
  40 & 1996/12/23 02:52 & 12/20 17:14 & 12/25 09:10 & 0.968 & 2.52+0.51-0.43 \\
  41 & 1996/12/28 09:23 & 12/25 09:15 & 12/31 04:42 & 0.967 & 2.09+0.44-0.36 \\
  42 & 1997/01/03 03:46 & 12/31 04:46 & 01/05 14:25 & 0.967 & 2.27+0.47-0.40 \\
  43 & 1997/01/08 04:17 & 01/05 14:31 & 01/10 14:11 & 0.967 & 3.12+0.54-0.47 \\
  44 & 1997/01/12 02:32 & 01/10 14:15 & 01/15 01:14 & 0.967 & 2.12+0.58-0.47 \\
  45 & 1997/01/17 20:51 & 01/15 02:10 & 01/20 09:04 & 0.968 & 1.95+0.54-0.45 \\
  46 & 1997/01/23 05:35 & 01/20 16:50 & 01/25 13:44 & 0.969 & 2.53+0.53-0.45 \\
  47 & 1997/01/28 13:08 & 01/25 13:52 & 01/31 08:09 & 0.970 & 2.33+0.46-0.40 \\
  48 & 1997/02/02 21:04 & 01/31 08:58 & 02/05 09:26 & 0.971 & 2.85+0.50-0.43 \\
  49 & 1997/02/08 01:15 & 02/05 10:09 & 02/10 16:01 & 0.973 & 2.40+0.47-0.40 \\
  50 & 1997/02/12 23:31 & 02/10 16:04 & 02/15 03:25 & 0.975 & 2.45+0.55-0.46 \\
  51 & 1997/02/18 04:20 & 02/15 04:58 & 02/20 06:56 & 0.977 & 2.15+0.55-0.46 \\
  52 & 1997/02/22 16:11 & 02/20 08:22 & 02/25 03:02 & 0.979 & 2.27+0.53-0.45 \\
  53 & 1997/02/27 00:36 & 02/25 03:53 & 02/28 22:36 & 0.981 & 1.78+0.49-0.40 \\
  54 & 1997/03/03 02:52 & 02/28 22:39 & 03/05 08:23 & 0.983 & 2.19+0.50-0.42 \\
  55 & 1997/03/08 05:42 & 03/05 17:59 & 03/10 09:39 & 0.985 & 2.70+0.61-0.50 \\
  56 & 1997/03/13 04:23 & 03/10 09:46 & 03/15 22:46 & 0.988 & 2.37+0.45-0.38 \\
  57 & 1997/03/18 00:12 & 03/15 22:49 & 03/20 02:16 & 0.990 & 2.51+0.56-0.47 \\
  58 & 1997/03/22 12:30 & 03/20 02:26 & 03/25 08:29 & 0.993 & 2.03+0.49-0.41 \\
  59 & 1997/03/28 10:23 & 03/25 08:33 & 03/31 09:21 & 0.996 & 2.03+0.43-0.36 \\
  60 & 1997/04/03 04:23 & 03/31 21:11 & 04/05 13:02 & 1.000 & 2.63+0.52-0.45 \\
\hline
\end{tabular}
\begin{tabular}{|r|c|c|c|c|c|}
\hline
No. & t$_w$ & t$_s$ & t$_e$ & $R^2$ & $\phi_{\nu}$ \\
\hline
  61 & 1997/04/07 20:00 & 04/05 13:22 & 04/10 03:03 & 1.002 & 1.65+0.47-0.38 \\
  62 & 1997/04/12 14:02 & 04/10 03:46 & 04/15 05:51 & 1.005 & 2.61+0.52-0.45 \\
  63 & 1997/04/18 02:57 & 04/15 06:13 & 04/20 04:41 & 1.008 & 2.21+0.57-0.48 \\
  64 & 1997/04/21 23:13 & 04/20 07:17 & 04/25 00:07 & 1.010 & 3.20+0.83-0.67 \\
  65 & 1997/04/28 07:51 & 04/25 01:13 & 04/30 00:22 & 1.014 & 2.26+0.68-0.54 \\
  66 & 1997/05/03 00:50 & 04/30 00:31 & 05/05 15:37 & 1.016 & 2.59+0.50-0.43 \\
  67 & 1997/05/07 23:46 & 05/05 15:39 & 05/10 03:37 & 1.018 & 1.69+0.48-0.39 \\
  68 & 1997/05/12 19:48 & 05/10 04:19 & 05/15 09:10 & 1.021 & 2.86+0.51-0.44 \\
  69 & 1997/05/17 20:55 & 05/15 09:12 & 05/20 00:38 & 1.023 & 2.05+0.50-0.41 \\
  70 & 1997/05/22 19:55 & 05/20 00:41 & 05/25 09:09 & 1.025 & 2.14+0.55-0.45 \\
  71 & 1997/05/27 20:45 & 05/25 09:12 & 05/31 05:47 & 1.027 & 2.25+0.53-0.43 \\
  72 & 1997/06/03 13:01 & 05/31 06:10 & 06/05 04:30 & 1.029 & 1.82+0.92-0.63 \\
  73 & 1997/06/07 15:22 & 06/05 05:52 & 06/10 02:33 & 1.030 & 2.69+0.53-0.46 \\
  74 & 1997/06/12 13:28 & 06/10 03:33 & 06/15 05:11 & 1.031 & 1.69+0.47-0.39 \\
  75 & 1997/06/17 17:29 & 06/15 05:37 & 06/20 12:17 & 1.032 & 2.72+0.51-0.44 \\
  76 & 1997/06/22 18:27 & 06/20 12:22 & 06/25 07:47 & 1.033 & 2.54+0.50-0.44 \\
  77 & 1997/06/28 08:00 & 06/25 07:51 & 06/30 02:28 & 1.033 & 2.89+0.61-0.52 \\
  78 & 1997/07/02 17:57 & 06/30 02:31 & 07/05 01:17 & 1.034 & 1.84+0.48-0.41 \\
  79 & 1997/07/07 19:24 & 07/05 01:53 & 07/10 07:24 & 1.034 & 2.07+0.46-0.40 \\
  80 & 1997/07/12 21:04 & 07/10 07:29 & 07/15 10:22 & 1.033 & 1.34+0.41-0.35 \\
  81 & 1997/07/17 22:57 & 07/15 10:25 & 07/20 12:47 & 1.033 & 3.11+0.51-0.45 \\
  82 & 1997/07/22 11:16 & 07/20 12:55 & 07/25 04:51 & 1.032 & 1.94+0.57-0.48 \\
  83 & 1997/07/28 13:06 & 07/25 05:18 & 07/31 00:40 & 1.031 & 2.83+0.51-0.44 \\
  84 & 1997/08/02 13:56 & 07/31 01:04 & 08/05 00:35 & 1.030 & 2.28+0.45-0.38 \\
  85 & 1997/08/07 15:44 & 08/05 00:39 & 08/10 05:33 & 1.028 & 2.34+0.46-0.40 \\
  86 & 1997/08/12 16:50 & 08/10 05:38 & 08/15 06:15 & 1.027 & 2.43+0.56-0.48 \\
  87 & 1997/08/17 21:18 & 08/15 06:19 & 08/20 14:35 & 1.025 & 1.79+0.42-0.36 \\
  88 & 1997/08/23 06:06 & 08/20 14:45 & 08/25 20:02 & 1.023 & 2.25+0.46-0.41 \\
  89 & 1997/08/28 17:26 & 08/25 20:03 & 08/31 13:09 & 1.020 & 2.68+0.44-0.39 \\
  90 & 1997/09/02 23:59 & 08/31 13:11 & 09/05 15:30 & 1.018 & 1.34+0.47-0.40 \\
  91 & 1997/09/07 05:35 & 09/05 15:32 & 09/10 04:48 & 1.016 & 2.11+0.60-0.49 \\
  92 & 1997/09/14 05:31 & 09/10 04:50 & 09/15 21:50 & 1.012 & 2.12+0.61-0.51 \\
  93 & 1997/10/04 14:39 & 09/30 00:42 & 10/05 10:51 & 1.001 & 2.71+0.89-0.70 \\
  94 & 1997/10/06 22:10 & 10/05 11:11 & 10/10 18:48 & 0.999 & 2.25+0.71-0.55 \\
  95 & 1997/10/13 10:04 & 10/10 18:50 & 10/15 17:15 & 0.995 & 2.01+0.53-0.44 \\
  96 & 1997/10/18 13:13 & 10/15 18:52 & 10/20 16:16 & 0.993 & 2.06+0.70-0.56 \\
  97 & 1997/10/22 14:08 & 10/20 18:13 & 10/25 17:55 & 0.990 & 4.26+0.98-0.81 \\
  98 & 1997/10/28 04:09 & 10/25 17:57 & 10/30 14:40 & 0.987 & 2.54+0.46-0.40 \\
  99 & 1997/11/02 13:49 & 10/30 14:49 & 11/05 14:02 & 0.984 & 1.74+0.39-0.33 \\
 100 & 1997/11/08 06:12 & 11/05 14:04 & 11/10 18:55 & 0.982 & 1.50+0.38-0.32 \\
 101 & 1997/11/13 08:32 & 11/11 00:53 & 11/15 09:59 & 0.979 & 2.66+0.50-0.43 \\
 102 & 1997/11/18 03:46 & 11/15 13:26 & 11/20 19:59 & 0.977 & 2.91+0.54-0.46 \\
 103 & 1997/11/22 13:56 & 11/20 20:02 & 11/25 00:20 & 0.975 & 3.06+0.60-0.52 \\
 104 & 1997/11/27 05:32 & 11/25 00:50 & 11/30 02:45 & 0.974 & 2.23+0.51-0.43 \\
 105 & 1997/12/02 10:36 & 11/30 04:04 & 12/05 10:18 & 0.972 & 3.02+0.60-0.52 \\
 106 & 1997/12/07 23:43 & 12/05 10:35 & 12/10 11:35 & 0.970 & 2.30+0.47-0.41 \\
 107 & 1997/12/12 20:02 & 12/10 11:38 & 12/15 05:25 & 0.969 & 2.31+0.44-0.38 \\
 108 & 1997/12/17 22:06 & 12/15 05:27 & 12/20 13:00 & 0.968 & 2.30+0.47-0.40 \\
 109 & 1997/12/23 02:06 & 12/20 13:14 & 12/25 16:12 & 0.968 & 2.65+0.46-0.40 \\
 110 & 1997/12/28 21:09 & 12/25 16:15 & 12/31 21:10 & 0.967 & 2.46+0.41-0.36 \\
 111 & 1998/01/02 13:57 & 12/31 21:12 & 01/05 10:10 & 0.968 & 2.13+0.55-0.46 \\
 112 & 1998/01/08 05:37 & 01/05 12:39 & 01/10 23:32 & 0.967 & 2.72+0.51-0.44 \\
 113 & 1998/01/13 14:01 & 01/10 23:34 & 01/16 00:27 & 0.967 & 2.16+0.55-0.48 \\
 114 & 1998/01/18 08:12 & 01/16 01:00 & 01/20 08:45 & 0.968 & 2.60+0.54-0.46 \\
 115 & 1998/01/23 02:57 & 01/20 09:11 & 01/25 18:05 & 0.969 & 2.37+0.44-0.39 \\
 116 & 1998/01/28 14:49 & 01/25 18:20 & 01/31 11:16 & 0.970 & 2.20+0.40-0.35 \\
 117 & 1998/02/02 20:45 & 01/31 11:18 & 02/05 02:46 & 0.971 & 1.60+0.43-0.36 \\
 118 & 1998/02/07 17:38 & 02/05 03:14 & 02/10 06:59 & 0.973 & 1.63+0.41-0.34 \\
 119 & 1998/02/12 23:28 & 02/10 07:02 & 02/15 08:26 & 0.975 & 2.57+0.47-0.41 \\
 120 & 1998/02/17 17:14 & 02/15 08:28 & 02/20 08:10 & 0.976 & 2.22+0.45-0.39 \\
\hline
\end{tabular}
\end{table}
\begin{table}[!ht]
\tiny
\begin{tabular}{|r|c|c|c|c|c|}
\hline
No. & t$_w$ & t$_s$ & t$_e$ & $R^2$ & $\phi_{\nu}$ \\
\hline
 121 & 1998/02/23 08:39 & 02/20 09:45 & 02/25 17:10 & 0.979 & 2.66+0.48-0.42 \\
 122 & 1998/02/27 05:25 & 02/25 17:12 & 02/28 17:32 & 0.981 & 1.80+0.53-0.43 \\
 123 & 1998/03/02 22:08 & 02/28 17:35 & 03/05 10:44 & 0.982 & 1.75+0.46-0.39 \\
 124 & 1998/03/08 00:22 & 03/05 12:10 & 03/10 06:37 & 0.985 & 2.39+0.48-0.42 \\
 125 & 1998/03/12 13:34 & 03/10 06:41 & 03/15 06:08 & 0.987 & 2.14+0.45-0.39 \\
 126 & 1998/03/18 12:54 & 03/15 07:09 & 03/20 00:39 & 0.991 & 2.85+0.62-0.52 \\
 127 & 1998/03/22 23:02 & 03/20 00:42 & 03/25 21:56 & 0.993 & 2.32+0.40-0.35 \\
 128 & 1998/03/28 19:09 & 03/25 21:58 & 03/31 15:49 & 0.996 & 2.24+0.39-0.35 \\
 129 & 1998/04/03 02:58 & 03/31 15:51 & 04/05 13:32 & 0.999 & 2.20+0.43-0.38 \\
 130 & 1998/04/08 06:34 & 04/05 13:35 & 04/10 22:50 & 1.002 & 1.61+0.38-0.32 \\
 131 & 1998/04/13 05:22 & 04/10 22:52 & 04/15 23:13 & 1.005 & 2.84+0.52-0.45 \\
 132 & 1998/04/18 02:26 & 04/15 23:15 & 04/20 10:41 & 1.008 & 1.68+0.44-0.37 \\
 133 & 1998/04/23 00:08 & 04/20 11:35 & 04/25 11:29 & 1.011 & 3.51+0.51-0.45 \\
 134 & 1998/04/27 22:58 & 04/25 11:32 & 04/30 10:27 & 1.013 & 2.82+0.46-0.40 \\
 135 & 1998/05/03 00:32 & 04/30 10:48 & 05/05 14:04 & 1.016 & 2.64+0.44-0.39 \\
 136 & 1998/05/07 22:19 & 05/05 14:06 & 05/10 06:12 & 1.018 & 2.22+0.46-0.39 \\
 137 & 1998/05/12 18:28 & 05/10 06:14 & 05/15 10:05 & 1.021 & 1.93+0.46-0.39 \\
 138 & 1998/05/18 04:13 & 05/15 20:20 & 05/20 12:03 & 1.023 & 1.85+0.44-0.37 \\
 139 & 1998/05/23 00:59 & 05/20 12:06 & 05/25 10:06 & 1.025 & 2.17+0.45-0.39 \\
 140 & 1998/05/28 13:15 & 05/25 10:08 & 05/31 16:14 & 1.027 & 2.58+0.42-0.37 \\
 141 & 1998/06/02 20:08 & 05/31 16:15 & 06/05 00:44 & 1.029 & 2.58+0.50-0.44 \\
 142 & 1998/06/07 15:53 & 06/05 00:46 & 06/10 07:54 & 1.030 & 2.65+0.45-0.40 \\
 143 & 1998/06/12 16:53 & 06/10 07:57 & 06/15 00:50 & 1.031 & 2.09+0.44-0.37 \\
 144 & 1998/06/17 17:56 & 06/15 00:52 & 06/20 11:44 & 1.032 & 2.76+0.45-0.39 \\
 145 & 1998/06/22 21:10 & 06/20 11:48 & 06/25 09:24 & 1.033 & 2.19+0.45-0.39 \\
 146 & 1998/06/28 03:33 & 06/25 20:56 & 06/30 00:00 & 1.033 & 2.37+0.51-0.43 \\
 147 & 1998/07/03 03:47 & 06/30 00:02 & 07/05 22:38 & 1.034 & 1.58+0.39-0.33 \\
 148 & 1998/07/08 02:01 & 07/05 22:40 & 07/10 23:08 & 1.034 & 3.01+0.55-0.47 \\
 149 & 1998/07/12 20:57 & 07/10 23:13 & 07/15 00:04 & 1.033 & 1.65+0.48-0.41 \\
 150 & 1998/07/17 23:19 & 07/15 00:08 & 07/20 13:25 & 1.033 & 2.23+0.42-0.37 \\
 151 & 1998/07/22 19:15 & 07/20 13:26 & 07/25 00:18 & 1.032 & 2.62+0.47-0.41 \\
 152 & 1998/07/28 08:41 & 07/25 00:19 & 07/31 19:08 & 1.031 & 1.88+0.34-0.30 \\
 153 & 1998/08/03 00:04 & 07/31 19:13 & 08/05 04:38 & 1.030 & 1.50+0.40-0.33 \\
 154 & 1998/08/08 00:02 & 08/05 04:40 & 08/10 13:48 & 1.028 & 2.39+0.42-0.37 \\
 155 & 1998/08/13 05:13 & 08/10 13:51 & 08/15 18:27 & 1.027 & 1.80+0.42-0.36 \\
 156 & 1998/08/18 05:19 & 08/15 18:28 & 08/20 13:30 & 1.025 & 3.06+0.48-0.43 \\
 157 & 1998/08/23 07:08 & 08/20 13:32 & 08/25 21:10 & 1.023 & 2.02+0.40-0.34 \\
 158 & 1998/08/28 05:42 & 08/25 21:36 & 08/31 08:08 & 1.021 & 3.65+0.53-0.47 \\
 159 & 1998/09/02 23:09 & 08/31 08:29 & 09/05 11:36 & 1.018 & 2.71+0.45-0.39 \\
 160 & 1998/09/08 05:11 & 09/05 11:38 & 09/10 15:07 & 1.015 & 2.83+0.47-0.41 \\
 161 & 1998/09/13 05:34 & 09/10 15:17 & 09/15 13:30 & 1.013 & 1.92+0.41-0.35 \\
 162 & 1998/09/18 01:01 & 09/15 13:31 & 09/20 15:19 & 1.010 & 3.16+0.47-0.42 \\
 163 & 1998/09/22 17:44 & 09/20 15:20 & 09/25 00:05 & 1.007 & 2.13+0.45-0.39 \\
 164 & 1998/09/27 14:46 & 09/25 00:27 & 09/30 05:05 & 1.005 & 1.96+0.40-0.35 \\
 165 & 1998/10/02 04:12 & 09/30 05:10 & 10/05 01:57 & 1.002 & 2.20+0.51-0.43 \\
 166 & 1998/10/07 21:23 & 10/05 02:00 & 10/10 15:33 & 0.999 & 2.22+0.40-0.35 \\
 167 & 1998/10/12 20:22 & 10/10 15:35 & 10/15 00:57 & 0.996 & 2.97+0.50-0.44 \\
 168 & 1998/10/17 22:14 & 10/15 01:01 & 10/20 20:36 & 0.993 & 2.31+0.38-0.34 \\
 169 & 1998/10/23 11:10 & 10/20 20:39 & 10/25 12:40 & 0.990 & 1.78+0.46-0.38 \\
 170 & 1998/10/27 22:50 & 10/25 12:44 & 10/30 07:51 & 0.988 & 2.17+0.42-0.36 \\
 171 & 1998/11/02 03:57 & 10/30 07:54 & 11/05 09:42 & 0.985 & 2.42+0.41-0.36 \\
 172 & 1998/11/09 04:13 & 11/05 09:46 & 11/10 15:13 & 0.981 & 3.35+0.82-0.69 \\
 173 & 1998/11/13 10:37 & 11/10 18:00 & 11/16 03:41 & 0.979 & 2.84+0.67-0.56 \\
 174 & 1998/11/18 16:24 & 11/16 03:44 & 11/20 13:50 & 0.977 & 3.04+0.58-0.51 \\
 175 & 1998/11/21 14:21 & 11/20 13:52 & 11/25 13:31 & 0.976 & 2.73+1.19-0.90 \\
 176 & 1998/11/29 11:49 & 11/25 14:10 & 11/30 08:38 & 0.973 & 2.37+0.77-0.61 \\
 177 & 1998/12/04 03:26 & 11/30 09:28 & 12/05 22:18 & 0.972 & 2.94+0.62-0.53 \\
 178 & 1998/12/08 05:22 & 12/05 22:22 & 12/10 12:26 & 0.970 & 2.96+0.49-0.43 \\
 179 & 1998/12/12 03:41 & 12/10 12:31 & 12/15 01:02 & 0.969 & 2.39+0.56-0.46 \\
 180 & 1998/12/18 13:02 & 12/15 01:04 & 12/20 05:27 & 0.968 & 2.41+0.69-0.57 \\
\hline
\end{tabular}
\begin{tabular}{|r|c|c|c|c|c|}
\hline
No. & t$_w$ & t$_s$ & t$_e$ & $R^2$ & $\phi_{\nu}$ \\
\hline
 181 & 1998/12/22 16:05 & 12/20 05:31 & 12/25 02:27 & 0.968 & 3.36+0.60-0.52 \\
 182 & 1998/12/28 12:01 & 12/25 03:33 & 12/31 13:01 & 0.967 & 2.31+0.42-0.37 \\
 183 & 1999/01/03 01:15 & 12/31 13:58 & 01/05 10:50 & 0.967 & 2.48+0.49-0.42 \\
 184 & 1999/01/08 11:13 & 01/06 01:11 & 01/10 20:01 & 0.967 & 3.16+0.63-0.54 \\
 185 & 1999/01/13 04:24 & 01/10 20:03 & 01/15 15:41 & 0.967 & 3.21+0.63-0.54 \\
 186 & 1999/01/17 19:04 & 01/15 15:43 & 01/20 00:03 & 0.968 & 2.37+0.48-0.41 \\
 187 & 1999/01/23 01:03 & 01/20 00:45 & 01/25 12:23 & 0.969 & 2.56+0.48-0.42 \\
 188 & 1999/01/28 14:39 & 01/25 12:25 & 01/31 18:46 & 0.970 & 2.81+0.42-0.38 \\
 189 & 1999/02/03 00:05 & 01/31 18:47 & 02/05 04:43 & 0.971 & 3.60+0.51-0.45 \\
 190 & 1999/02/07 18:40 & 02/05 04:44 & 02/10 09:15 & 0.973 & 2.27+0.49-0.42 \\
 191 & 1999/02/12 19:55 & 02/10 16:30 & 02/15 00:38 & 0.974 & 2.33+0.47-0.41 \\
 192 & 1999/02/17 15:08 & 02/15 00:40 & 02/20 00:00 & 0.976 & 3.09+0.53-0.46 \\
 193 & 1999/02/22 23:03 & 02/20 00:03 & 02/25 19:22 & 0.979 & 2.26+0.44-0.39 \\
 194 & 1999/02/26 23:42 & 02/25 20:57 & 02/28 02:13 & 0.980 & 2.80+0.72-0.58 \\
 195 & 1999/03/02 15:37 & 02/28 02:16 & 03/05 04:59 & 0.982 & 2.98+0.47-0.41 \\
 196 & 1999/03/07 10:04 & 03/05 05:02 & 03/10 00:33 & 0.985 & 2.40+0.49-0.42 \\
 197 & 1999/03/12 12:00 & 03/10 01:03 & 03/15 00:41 & 0.987 & 2.64+0.45-0.39 \\
 198 & 1999/03/17 17:26 & 03/15 01:21 & 03/20 08:31 & 0.990 & 1.79+0.40-0.34 \\
 199 & 1999/03/23 08:01 & 03/20 08:34 & 03/25 23:04 & 0.993 & 1.97+0.44-0.38 \\
 200 & 1999/03/28 20:48 & 03/25 23:05 & 03/31 16:11 & 0.996 & 2.40+0.48-0.41 \\
 201 & 1999/04/03 07:55 & 04/01 01:42 & 04/05 08:23 & 0.999 & 1.73+0.51-0.42 \\
 202 & 1999/04/08 01:01 & 04/05 08:31 & 04/10 00:02 & 1.002 & 2.70+0.55-0.47 \\
 203 & 1999/04/12 05:55 & 04/10 00:22 & 04/15 13:14 & 1.004 & 2.74+0.50-0.44 \\
 204 & 1999/04/29 15:02 & 04/25 03:57 & 04/30 19:59 & 1.014 & 1.90+0.66-0.54 \\
 205 & 1999/05/03 02:23 & 04/30 20:00 & 05/05 10:16 & 1.016 & 2.66+0.49-0.42 \\
 206 & 1999/05/07 19:17 & 05/05 10:19 & 05/10 04:14 & 1.018 & 2.36+0.46-0.39 \\
 207 & 1999/05/13 13:56 & 05/10 04:15 & 05/15 23:11 & 1.021 & 3.09+0.53-0.46 \\
 208 & 1999/05/18 15:42 & 05/15 23:13 & 05/20 16:02 & 1.023 & 2.20+0.55-0.46 \\
 209 & 1999/05/23 02:24 & 05/20 16:06 & 05/25 12:54 & 1.025 & 1.62+0.39-0.33 \\
 210 & 1999/05/28 08:23 & 05/25 12:55 & 05/31 04:51 & 1.027 & 2.46+0.46-0.40 \\
 211 & 1999/06/02 13:42 & 05/31 04:53 & 06/05 00:30 & 1.028 & 2.42+0.45-0.39 \\
 212 & 1999/06/07 22:21 & 06/05 00:36 & 06/10 18:57 & 1.030 & 2.40+0.41-0.36 \\
 213 & 1999/06/12 23:05 & 06/10 18:59 & 06/15 04:15 & 1.031 & 1.34+0.41-0.34 \\
 214 & 1999/06/17 11:06 & 06/15 04:17 & 06/20 02:39 & 1.032 & 2.19+0.50-0.42 \\
 215 & 1999/06/23 06:29 & 06/20 02:43 & 06/25 15:30 & 1.033 & 2.24+0.43-0.37 \\
 216 & 1999/06/27 23:37 & 06/25 15:31 & 06/30 05:21 & 1.033 & 2.63+0.47-0.41 \\
 217 & 1999/07/03 02:31 & 06/30 05:39 & 07/05 16:48 & 1.034 & 2.39+0.43-0.38 \\
 218 & 1999/07/08 07:33 & 07/05 17:32 & 07/10 08:12 & 1.034 & 2.11+0.47-0.40 \\
 219 & 1999/07/11 21:19 & 07/10 08:18 & 07/15 07:11 & 1.034 & 1.32+0.62-0.51 \\
 220 & 1999/07/18 11:06 & 07/15 13:26 & 07/20 00:09 & 1.033 & 2.19+0.76-0.60 \\
 221 & 1999/07/24 00:13 & 07/20 00:11 & 07/25 21:16 & 1.032 & 1.90+0.51-0.42 \\
 222 & 1999/07/28 18:54 & 07/25 21:20 & 07/31 02:30 & 1.031 & 1.87+0.48-0.40 \\
 223 & 1999/08/02 21:36 & 07/31 02:39 & 08/05 09:56 & 1.030 & 2.40+0.48-0.41 \\
 224 & 1999/08/08 01:16 & 08/05 12:09 & 08/10 15:46 & 1.028 & 1.54+0.42-0.36 \\
 225 & 1999/08/12 23:29 & 08/10 15:48 & 08/15 05:54 & 1.027 & 3.24+0.51-0.45 \\
 226 & 1999/08/17 23:12 & 08/15 05:55 & 08/20 17:18 & 1.025 & 2.59+0.45-0.40 \\
 227 & 1999/08/23 03:41 & 08/20 17:19 & 08/25 14:11 & 1.023 & 1.86+0.43-0.38 \\
 228 & 1999/08/28 19:28 & 08/25 14:18 & 08/31 11:54 & 1.020 & 2.78+0.47-0.41 \\
 229 & 1999/09/03 02:15 & 08/31 11:57 & 09/05 15:00 & 1.018 & 2.13+0.43-0.37 \\
 230 & 1999/09/08 01:13 & 09/05 15:01 & 09/10 13:27 & 1.015 & 1.74+0.42-0.35 \\
 231 & 1999/09/12 21:50 & 09/10 13:28 & 09/15 02:10 & 1.013 & 2.62+0.49-0.43 \\
 232 & 1999/09/18 01:38 & 09/15 02:18 & 09/20 21:40 & 1.010 & 2.64+0.44-0.39 \\
 233 & 1999/09/22 23:03 & 09/20 21:41 & 09/25 08:13 & 1.007 & 1.77+0.47-0.40 \\
 234 & 1999/09/28 02:39 & 09/25 08:36 & 09/30 20:40 & 1.005 & 2.54+0.40-0.35 \\
 235 & 1999/10/03 02:49 & 09/30 20:42 & 10/05 09:22 & 1.002 & 2.97+0.50-0.44 \\
 236 & 1999/10/07 21:29 & 10/05 17:25 & 10/10 12:04 & 0.999 & 1.47+0.42-0.35 \\
 237 & 1999/10/12 19:14 & 10/10 12:09 & 10/15 08:34 & 0.996 & 2.38+0.47-0.41 \\
 238 & 1999/10/18 06:27 & 10/15 14:33 & 10/21 00:12 & 0.993 & 2.16+0.44-0.38 \\
 239 & 1999/10/22 05:51 & 10/21 00:14 & 10/25 21:14 & 0.991 & 2.10+0.72-0.60 \\
 240 & 1999/10/28 12:18 & 10/25 23:33 & 10/30 10:26 & 0.987 & 1.61+0.60-0.50 \\
\hline
\end{tabular}
\end{table}
\begin{table}[!ht]
\tiny
\begin{tabular}{|r|c|c|c|c|c|}
\hline
No. & t$_w$ & t$_s$ & t$_e$ & $R^2$ & $\phi_{\nu}$ \\
\hline
 241 & 1999/11/02 09:00 & 10/30 10:28 & 11/05 08:27 & 0.985 & 2.68+0.41-0.36 \\
 242 & 1999/11/07 23:39 & 11/05 08:29 & 11/10 14:56 & 0.982 & 2.30+0.43-0.38 \\
 243 & 1999/11/13 04:01 & 11/10 14:57 & 11/15 11:58 & 0.980 & 1.91+0.46-0.40 \\
 244 & 1999/11/17 22:41 & 11/15 12:05 & 11/20 09:14 & 0.978 & 2.05+0.44-0.38 \\
 245 & 1999/11/22 21:23 & 11/20 09:15 & 11/25 08:58 & 0.976 & 2.06+0.42-0.36 \\
 246 & 1999/11/27 18:11 & 11/25 09:00 & 11/30 12:44 & 0.974 & 2.75+0.47-0.42 \\
 247 & 1999/12/04 10:20 & 11/30 18:24 & 12/05 23:02 & 0.972 & 2.57+0.58-0.50 \\
 248 & 1999/12/08 03:41 & 12/05 23:03 & 12/10 03:53 & 0.971 & 2.07+0.48-0.42 \\
 249 & 1999/12/12 22:41 & 12/10 08:07 & 12/15 14:25 & 0.969 & 1.97+0.41-0.35 \\
 250 & 1999/12/17 20:22 & 12/15 14:26 & 12/20 09:21 & 0.968 & 2.25+0.49-0.42 \\
 251 & 1999/12/24 00:01 & 12/20 10:20 & 12/25 10:01 & 0.968 & 2.34+0.54-0.47 \\
 252 & 1999/12/28 13:08 & 12/25 10:03 & 12/31 20:06 & 0.967 & 2.23+0.38-0.34 \\
 253 & 2000/01/02 22:26 & 12/31 20:07 & 01/05 01:29 & 0.967 & 2.69+0.50-0.44 \\
 254 & 2000/01/07 13:14 & 01/05 01:30 & 01/10 01:56 & 0.967 & 3.01+0.46-0.41 \\
 255 & 2000/01/12 23:31 & 01/10 01:58 & 01/15 18:05 & 0.967 & 2.33+0.40-0.36 \\
 256 & 2000/01/18 05:01 & 01/15 18:06 & 01/20 14:16 & 0.968 & 2.39+0.45-0.39 \\
 257 & 2000/01/23 01:12 & 01/20 14:33 & 01/25 11:20 & 0.969 & 2.36+0.45-0.40 \\
 258 & 2000/01/27 18:39 & 01/25 11:21 & 01/31 01:28 & 0.970 & 1.76+0.43-0.37 \\
 259 & 2000/02/02 21:41 & 01/31 01:39 & 02/05 15:25 & 0.971 & 2.57+0.41-0.36 \\
 260 & 2000/02/08 00:04 & 02/05 15:27 & 02/10 08:41 & 0.973 & 2.89+0.48-0.42 \\
 261 & 2000/02/12 21:13 & 02/10 08:42 & 02/15 09:29 & 0.974 & 2.32+0.45-0.39 \\
 262 & 2000/02/17 21:46 & 02/15 09:30 & 02/20 10:48 & 0.976 & 3.09+0.46-0.40 \\
 263 & 2000/02/23 00:12 & 02/20 10:49 & 02/25 14:13 & 0.978 & 2.38+0.43-0.37 \\
 264 & 2000/02/26 19:49 & 02/25 14:21 & 02/28 01:22 & 0.980 & 2.88+0.69-0.58 \\
 265 & 2000/03/02 07:38 & 02/28 01:23 & 03/05 10:11 & 0.982 & 1.97+0.37-0.32 \\
 266 & 2000/03/07 22:46 & 03/05 10:14 & 03/10 08:51 & 0.985 & 2.65+0.46-0.41 \\
 267 & 2000/03/12 17:34 & 03/10 08:55 & 03/15 01:59 & 0.988 & 3.20+0.50-0.44 \\
 268 & 2000/03/17 14:02 & 03/15 02:01 & 03/20 02:14 & 0.990 & 2.29+0.46-0.40 \\
 269 & 2000/03/22 20:26 & 03/20 02:23 & 03/25 13:16 & 0.993 & 2.34+0.44-0.38 \\
 270 & 2000/03/28 15:40 & 03/25 18:03 & 03/31 09:33 & 0.997 & 2.05+0.42-0.37 \\
 271 & 2000/04/03 08:43 & 03/31 20:08 & 04/05 15:27 & 1.000 & 1.75+0.45-0.38 \\
 272 & 2000/04/08 02:55 & 04/05 15:28 & 04/10 14:37 & 1.003 & 1.99+0.45-0.39 \\
 273 & 2000/04/13 13:29 & 04/10 15:24 & 04/15 09:12 & 1.006 & 1.32+0.55-0.42 \\
 274 & 2000/04/17 22:38 & 04/15 09:15 & 04/20 08:57 & 1.008 & 2.10+0.48-0.40 \\
 275 & 2000/04/23 02:15 & 04/20 14:20 & 04/25 00:13 & 1.011 & 1.97+0.54-0.45 \\
 276 & 2000/04/27 21:17 & 04/25 00:26 & 04/30 17:04 & 1.013 & 1.90+0.41-0.36 \\
 277 & 2000/05/03 03:06 & 04/30 17:20 & 05/05 08:52 & 1.016 & 2.14+0.46-0.39 \\
 278 & 2000/05/08 03:40 & 05/05 19:13 & 05/10 08:26 & 1.019 & 2.07+0.45-0.38 \\
 279 & 2000/05/13 04:59 & 05/10 08:28 & 05/15 16:00 & 1.021 & 2.15+0.43-0.37 \\
 280 & 2000/05/18 04:18 & 05/15 16:00 & 05/20 16:32 & 1.023 & 2.07+0.41-0.35 \\
 281 & 2000/05/23 02:46 & 05/20 16:33 & 05/25 14:33 & 1.025 & 2.66+0.49-0.43 \\
 282 & 2000/05/29 15:42 & 05/25 14:38 & 06/02 15:13 & 1.027 & 1.74+0.33-0.29 \\
 283 & 2000/06/08 00:31 & 06/02 15:14 & 06/11 13:08 & 1.030 & 2.17+0.57-0.47 \\
 284 & 2000/06/13 20:21 & 06/11 16:08 & 06/16 04:32 & 1.032 & 2.72+0.53-0.45 \\
 285 & 2000/06/18 07:12 & 06/16 04:33 & 06/20 08:52 & 1.032 & 1.72+0.44-0.38 \\
 286 & 2000/06/23 14:53 & 06/20 08:56 & 06/27 15:35 & 1.033 & 2.38+0.41-0.36 \\
 287 & 2000/06/29 01:52 & 06/27 15:36 & 06/30 17:36 & 1.034 & 3.01+0.71-0.59 \\
 288 & 2000/07/03 11:57 & 07/01 17:50 & 07/05 08:14 & 1.034 & 2.19+0.55-0.47 \\
 289 & 2000/07/09 12:57 & 07/05 08:14 & 07/13 14:47 & 1.034 & 1.89+0.34-0.30 \\
 290 & 2000/07/15 14:19 & 07/13 14:49 & 07/17 13:47 & 1.033 & 2.46+0.52-0.44 \\
 291 & 2000/07/21 12:41 & 07/17 14:50 & 07/25 08:09 & 1.032 & 2.17+0.36-0.32 \\
 292 & 2000/07/28 11:45 & 07/25 14:23 & 07/31 01:08 & 1.031 & 2.88+0.46-0.41 \\
 293 & 2000/08/03 12:26 & 07/31 01:10 & 08/06 22:33 & 1.030 & 2.13+0.39-0.34 \\
 294 & 2000/08/08 19:27 & 08/06 22:34 & 08/10 10:40 & 1.028 & 1.13+0.48-0.38 \\
 295 & 2000/08/14 21:56 & 08/10 10:47 & 08/19 18:16 & 1.026 & 2.13+0.34-0.30 \\
 296 & 2000/08/21 01:03 & 08/19 18:18 & 08/22 08:03 & 1.023 & 2.44+0.67-0.56 \\
 297 & 2000/08/23 20:25 & 08/22 08:11 & 08/25 12:40 & 1.022 & 1.82+0.66-0.55 \\
 298 & 2000/08/29 20:22 & 08/25 12:47 & 09/03 03:28 & 1.020 & 2.33+0.34-0.31 \\
 299 & 2000/09/05 01:40 & 09/03 03:30 & 09/06 23:38 & 1.017 & 2.77+0.55-0.48 \\
 300 & 2000/09/09 08:29 & 09/06 23:39 & 09/11 09:38 & 1.014 & 2.13+0.48-0.41 \\
\hline
\end{tabular}
\begin{tabular}{|r|c|c|c|c|c|}
\hline
No. & t$_w$ & t$_s$ & t$_e$ & $R^2$ & $\phi_{\nu}$ \\
\hline
 301 & 2000/09/13 22:38 & 09/11 15:22 & 09/16 11:07 & 1.012 & 1.88+0.45-0.38 \\
 302 & 2000/09/18 09:06 & 09/16 11:55 & 09/20 04:05 & 1.010 & 2.90+0.57-0.50 \\
 303 & 2000/09/22 20:15 & 09/20 04:09 & 09/25 08:41 & 1.007 & 2.42+0.46-0.40 \\
 304 & 2000/09/29 07:27 & 09/25 11:06 & 10/03 03:01 & 1.003 & 2.61+0.36-0.33 \\
 305 & 2000/10/04 15:19 & 10/03 03:02 & 10/06 14:25 & 1.000 & 2.22+0.57-0.46 \\
 306 & 2000/10/11 00:37 & 10/06 14:28 & 10/15 02:45 & 0.997 & 2.04+0.33-0.30 \\
 307 & 2000/10/16 19:14 & 10/15 02:49 & 10/18 14:45 & 0.993 & 2.08+0.52-0.43 \\
 308 & 2000/10/19 09:43 & 10/18 18:15 & 10/20 01:19 & 0.992 & 4.23+1.05-0.89 \\
 309 & 2000/10/23 08:29 & 10/20 01:22 & 10/26 09:55 & 0.990 & 2.04+0.40-0.35 \\
 310 & 2000/10/28 13:37 & 10/26 09:58 & 10/31 07:33 & 0.987 & 2.58+0.52-0.45 \\
 311 & 2000/11/03 08:47 & 10/31 07:34 & 11/06 09:32 & 0.984 & 2.43+0.43-0.37 \\
 312 & 2000/11/08 20:17 & 11/06 09:39 & 11/11 06:38 & 0.981 & 2.27+0.46-0.39 \\
 313 & 2000/11/14 05:34 & 11/11 07:11 & 11/16 21:41 & 0.979 & 2.25+0.44-0.38 \\
 314 & 2000/11/18 15:38 & 11/16 21:43 & 11/21 15:44 & 0.977 & 2.58+0.58-0.49 \\
 315 & 2000/11/24 15:49 & 11/21 17:26 & 11/27 11:32 & 0.975 & 2.54+0.42-0.37 \\
 316 & 2000/11/28 20:31 & 11/27 11:36 & 11/30 08:10 & 0.973 & 2.47+0.68-0.56 \\
 317 & 2000/12/03 07:02 & 11/30 08:26 & 12/06 10:49 & 0.972 & 3.01+0.55-0.47 \\
 318 & 2000/12/08 06:40 & 12/06 10:50 & 12/10 02:44 & 0.970 & 1.17+0.42-0.34 \\
 319 & 2000/12/15 11:16 & 12/10 02:46 & 12/19 16:55 & 0.969 & 2.46+0.35-0.32 \\
 320 & 2000/12/20 07:49 & 12/19 17:08 & 12/20 23:17 & 0.968 & 1.87+1.21-0.87 \\
 321 & 2000/12/24 06:28 & 12/20 23:32 & 12/27 18:25 & 0.967 & 2.89+0.43-0.38 \\
 322 & 2001/01/01 09:17 & 12/27 18:30 & 01/05 09:49 & 0.967 & 2.56+0.35-0.31 \\
 323 & 2001/01/06 04:44 & 01/05 09:52 & 01/06 22:30 & 0.967 & 1.70+0.83-0.61 \\
 324 & 2001/01/09 11:08 & 01/06 22:32 & 01/11 16:44 & 0.967 & 2.71+0.59-0.51 \\
 325 & 2001/01/13 22:26 & 01/11 16:46 & 01/16 14:26 & 0.967 & 1.25+0.40-0.34 \\
 326 & 2001/01/19 03:49 & 01/16 14:32 & 01/22 03:19 & 0.968 & 3.55+0.49-0.43 \\
 327 & 2001/01/24 22:11 & 01/22 03:21 & 01/27 12:17 & 0.969 & 2.65+0.44-0.39 \\
 328 & 2001/01/29 08:26 & 01/27 12:20 & 01/31 03:25 & 0.970 & 2.75+0.54-0.46 \\
 329 & 2001/02/02 23:50 & 01/31 03:28 & 02/05 21:23 & 0.971 & 2.74+0.44-0.40 \\
 330 & 2001/02/08 15:36 & 02/05 21:26 & 02/11 06:17 & 0.973 & 2.21+0.40-0.35 \\
 331 & 2001/02/14 07:28 & 02/11 06:19 & 02/16 23:11 & 0.975 & 2.74+0.50-0.44 \\
 332 & 2001/02/18 09:49 & 02/16 23:14 & 02/20 01:26 & 0.977 & 3.45+0.66-0.56 \\
 333 & 2001/02/23 16:10 & 02/20 01:35 & 02/26 09:02 & 0.979 & 1.62+0.42-0.36 \\
 334 & 2001/03/01 01:19 & 02/26 15:02 & 03/02 08:36 & 0.982 & 2.07+0.60-0.48 \\
 335 & 2001/03/04 09:05 & 03/02 08:38 & 03/06 08:26 & 0.983 & 2.29+0.48-0.41 \\
 336 & 2001/03/08 08:08 & 03/06 08:28 & 03/10 04:39 & 0.985 & 3.39+0.58-0.50 \\
 337 & 2001/03/15 08:39 & 03/10 04:44 & 03/23 09:21 & 0.989 & 2.44+0.29-0.27 \\
 338 & 2001/03/26 13:04 & 03/23 09:24 & 03/29 16:56 & 0.995 & 2.75+0.41-0.37 \\
 339 & 2001/03/31 21:34 & 03/29 18:53 & 04/03 00:02 & 0.998 & 2.18+0.47-0.41 \\
 340 & 2001/04/06 04:43 & 04/03 00:05 & 04/09 10:16 & 1.001 & 1.67+0.36-0.31 \\
 341 & 2001/04/12 20:47 & 04/10 14:50 & 04/15 07:42 & 1.005 & 3.20+0.52-0.46 \\
 342 & 2001/04/18 02:57 & 04/15 07:46 & 04/20 22:30 & 1.008 & 2.08+0.41-0.36 \\
 343 & 2001/04/24 11:57 & 04/20 22:33 & 04/28 08:53 & 1.012 & 2.70+0.40-0.36 \\
 344 & 2001/04/29 17:29 & 04/28 19:54 & 04/30 09:36 & 1.014 & 2.05+1.10-0.78 \\
 345 & 2001/05/03 05:49 & 04/30 22:18 & 05/05 10:11 & 1.016 & 3.23+0.67-0.57 \\
 346 & 2001/05/08 05:03 & 05/05 20:49 & 05/10 16:53 & 1.019 & 2.46+0.67-0.56 \\
 347 & 2001/05/13 19:11 & 05/10 16:55 & 05/16 10:09 & 1.021 & 2.62+0.48-0.42 \\
 348 & 2001/05/18 18:15 & 05/16 10:13 & 05/21 01:50 & 1.023 & 3.49+0.52-0.46 \\
 349 & 2001/05/24 19:08 & 05/21 01:52 & 05/28 05:22 & 1.026 & 2.95+0.41-0.37 \\
 350 & 2001/06/01 01:32 & 05/28 05:25 & 06/04 20:38 & 1.028 & 2.40+0.35-0.32 \\
 351 & 2001/06/05 14:10 & 06/04 20:41 & 06/06 07:34 & 1.030 & 2.69+0.90-0.70 \\
 352 & 2001/06/08 17:38 & 06/06 07:37 & 06/11 02:23 & 1.030 & 2.80+0.47-0.41 \\
 353 & 2001/06/13 08:58 & 06/11 02:26 & 06/15 14:03 & 1.031 & 1.84+0.43-0.37 \\
 354 & 2001/06/18 02:54 & 06/15 14:07 & 06/20 16:35 & 1.032 & 2.77+0.43-0.38 \\
 355 & 2001/06/23 13:36 & 06/20 16:38 & 06/26 16:58 & 1.033 & 1.60+0.38-0.33 \\
 356 & 2001/06/28 18:33 & 06/26 17:01 & 06/30 20:07 & 1.033 & 1.21+0.43-0.36 \\
 357 & 2001/07/05 14:02 & 06/30 20:08 & 07/10 11:38 & 1.034 & 2.44+0.31-0.28 \\
 358 & 2001/07/12 16:28 & 07/10 12:40 & 07/15 08:34 & 1.033 & 2.63+0.51-0.45 \\
     &                  &             &             &       &                \\
     &                  &             &             &       &                \\
\hline
\end{tabular}
\end{table}


\end{document}